\pdfoutput=1
\documentclass[aps,amsfonts,amsmath,prd,preprint,nofootinbib]{revtex4}
\newcommand{\beq}{\begin{equation}}
\newcommand{\eeq}{\end{equation}}

\usepackage{graphicx,subfigure}

\def\bea{\begin{eqnarray}}
\def\eea{\end{eqnarray}}

\begin{document}

\title{The Initial State of a Primordial Anisotropic Stage of Inflation}

\author{Jose J. Blanco-Pillado${}^{a,b}$ and Masato Minamitsuji${}^{c,d}$}
\affiliation{$^a$ Dept. of Theoretical Physics, University of the Basque Country, Bilbao, Spain}
\affiliation{$^b$ IKERBASQUE, Basque Foundation for Science, 48013, Bilbao, Spain}
\affiliation{$^c$YITP, Kyoto University, Kyoto 606-8502, Japan}
\affiliation{$^d$CENTRA, Instituto Superior Tecnico,
Universidade de Lisboa, Avenida Rovisco Pais 1, 1049-001, Portugal}
\vskip 0.5cm
\begin{abstract} 
\vskip 0.5cm
We investigate the possibility that the inflationary period in the early universe was preceded by a primordial stage of strong anisotropy. In particular we focus on the simplest model of this kind, where the spacetime is described by a non-singular Kasner solution that quickly evolves into an isotropic de Sitter space, the so-called Kasner-de Sitter solution. The initial Big Bang singularity is replaced, in this case, by a horizon. We show that the extension of this metric to the region behind the horizon contains a timelike singularity which will be visible by cosmological observers. This makes it impossible to have a reliable prediction of the quantum state of the cosmological perturbations in the region of interest.  In this paper we consider the possibility that this Kasner-de Sitter universe is obtained as a result of a quantum tunneling process effectively substituting the region behind the horizon by an anisotropic parent vacuum state, namely a  $1+1$ dimensional spacetime compactified over an internal flat torus, $T_2$, which we take it to be of the form  $\text{{\it de Sitter}}_2 \times T_2$ or ${\text {\it Minkowski}}_2 \times T_2$. As a first approximation to understand the effects of this anisotropic initial state, we compute the power  spectrum of a massless scalar field in these backgrounds. In both cases, the spectrum converges at small scales to the isotropic scale invariant form and only present important deviations from it at the largest possible scales.  We find that the decompactification scenario from $M_2 \times T_2$ leads to a suppressed and slightly anisotropic power spectrum at large scales which could be related to some of the anomalies present in the current CMB data. On the other hand, the spectrum of the universe with a $dS_2 \times T_2$ parent vacuum presents an enhancement in power at large scales not consistent with observations.

\end{abstract}
 
\maketitle
\thispagestyle{empty}
\section{Introduction}

The latest results from the WMAP and {\it Planck} collaborations fit beautifully within 
a very simple model of inflation \cite{Hinshaw:2012aka,Ade:2013uln}. On the other hand, there are a number
of intriguing large scale anomalies in the cosmic microwave background (CMB) data
that clearly deserve some attention. These anomalies include the low power of
the quadrupole \cite{Hinshaw:1996ut}, the alignment of the quadrupole and octopole \cite{Tegmark:2003ve,Schwarz:2004gk},
the oscillation in the power of low $\ell<10$ multiples with $P_{odd} < P_{even}$ \cite{Land:2005jq}, 
as well as the so-called dipolar modulation \cite{Eriksen:2003db, Akrami:2014eta}. Although the statistical 
significance of some of these effects is still under debate, it is particularly
interesting to think that they might be related. Here we explore the possibility
that they may be due to a particular state of the universe at the onset of inflation.

Several authors have investigated the idea that some of these anomalies could
be due to a period of anisotropic inflation 
\cite{Ackerman:2007nb, Watanabe:2009ct,Barrow:2009gx,Emami:2010rm,Bartolo:2013msa,Ohashi:2013mka}. These models require the existence 
of some kind of matter during inflation that sustains an anisotropic energy
momentum tensor and bypass the no-hair theorems for a spacetime
with a positive cosmological constant \cite{Wald:1983ky}. It is interesting to see that some of these
models lead to an attractor behavior for this anisotropic period making their
predictions more robust (See, for example \cite{Soda:2012zm}).

There is however another way to explain these large scale anomalies without invoking the
presence of new anisotropic energy sources. The idea is to assume that inflation 
only lasted for a relatively short number of e-foldings, in fact, just enough to 
solve the horizon, flatness and isotropy problems. In a situation like this
one could be seeing the effects of the initial state of inflation at the largest
possible scales of the CMB today. This obviously requires a degree of fine-tuning of the number
of e-foldings, but taking into account the number of suspicious effects at those
scales one is tempted to consider this possibility seriously. In particular
one would like to explain the apparent violation of rotational symmetry
at the largest scales by a initial period of anisotropic evolution. Considering
a universe dominated by a pure cosmological constant, one would find
(in agreement with the no-hair theorems) a rapid approach to isotropic 
expansion. In other words, there is only a {\it primordial anisotropic stage} of inflation. This 
is the kind of scenario we are contemplating in this paper.

Such period of primordial anisotropic inflation was first considered in 
\cite{Gumrukcuoglu:2007bx} and \cite{Pitrou:2008gk}. One of the crucial points of this scenario is that,
of course, one does not have a long period of inflation that would
settle the quantum state to a Bunch-Davis vacuum, as one has for
a regular inflationary model with a large number of e-folds. This makes
the initial state of the vacuum before inflation potentially observable.
Here we would like to explore this possibility in more detail by looking
at some of the simple models that have been proposed for primordial
anisotropic inflation.

The rest of the paper is organized as follows. We show in Section II that the models studied in the literature can be extended
past their apparent Big Bang hypersurface to a spacelike region that has a 
timelike singularity. Furthermore, the quantum state for perturbations in
these models could not be specified without some knowledge about
the conditions on the singularity. This makes it impossible to make 
precise predictions on these scenarios. We present in section III
a different scenario where we replace the region with the
singularity with a lower dimensional compactified spacetime. This gives a new 
interpretation to the background geometry for anisotropic inflation as an anisotropic bubble created by the
decompactification of the lower dimensional state and 
allows us to obtain a consistent quantum initial state for the cosmological perturbations in this model. 
We discuss the way to find this quantum state in Section IV. Finally in section V we study the observational
consequences of such model and demonstrate that  in some cases the new quantum state 
alleviates some of the pathologies found in the power spectrum of perturbations.
We end in section VI with some discussion and conclusions.

\setcounter{page}{1}

\setcounter{footnote}{0}

\section{The Background Anisotropic Geometry}

\subsection{A Cosmological Background}

In this paper, we consider the possibility that
the geometry of our universe is described by the Bianchi I metric of the form,
\bea
ds^2= -dt^2 + \sum_{i=1}^3 a_i(t)^2dx_i^2,
\eea
where $a_i(t)$ ($i=1,2,3$) are the scale factors in the three different spatial directions. 
If the existence of matter is ignored in the primordial stage of the universe,
the initial metric can be approximated by the Kasner solution,
\bea
\label{kasner}
ds^2= -dt^2 + \sum_{i=1}^3 t^{2p_i}dx_i^2,
\eea
the vacuum solution of Einstein's equations,
where the three exponents satisfy the relations 
\bea
\label{p-conditions}
\sum_i p_i= \sum_i p_i^2=1.
\eea
In the presence of a positive cosmological constant $\Lambda>0$
(or the equivalent potential energy),
one can find solutions of Einstein's equations whose
geometries interpolate between the initial Kasner \eqref{kasner} solution
and a late-time isotropic de Sitter phase. These solutions are
given by the so-called Kasner-de Sitter solution
\cite{Saunders, Ellis, Sato, Hervik:2000ed, Gumrukcuoglu:2007bx,Pitrou:2008gk,Kim:2012bv,Dey:2013tfa,Dey:2012qp,Dey:2011mj},
\bea
\label{kasner-ds}
ds^2= -dt^2 + \sum_{i=1}^3
\sinh^{\frac{2}{3}} (3Ht)
\Big\{\tanh\Big(\frac{3Ht}{2}\Big)\Big\}^{2(p_i-\frac{1}{3})}
dx_i^2,
\eea
where $H:=\sqrt{\frac{\Lambda}{3}}$ is the Hubble rate of the de Sitter
in the late-time limit and the $p_i$ parameters satisfy the same conditions
as before, namely, Eq. (\ref{p-conditions}).

These spacetimes are initially anisotropic but this phase is short lived and within a period of
$t\sim ({\rm a\,\,\, few})\times  H^{-1}$ the universe becomes isotropic in agreement with the expectation of the cosmic no-hair theorem
\cite{Wald:1983ky}. The curvature invariant $R^{\mu\nu\rho\sigma}R_{\mu\nu\rho\sigma}$ at $t=0$ diverges
for a generic Kasner spacetime $\eqref{kasner}$, and thus the geometry is initially singular
except for the particular branch where  $p_1=1$ and $p_2=p_3=0$ of the Kasner-de Sitter solution 
\eqref{kasner-ds} that takes the form
\begin{equation}
ds^2= -dt^2 + \left({2\over 3} H^{-1} \sinh{{ 3 H t}\over {2 }} \left(\cosh {{3 H t }\over {2 }}\right)^{-{1\over 3}}\right)^2 dr^2 %dx^2
+ \left(\cosh {{3 H t}\over {2}}\right)^{4\over 3} %(dy^2 + dz^2)
dx_\perp^2~,
\label{Kasner-deSitter-1}
\end{equation}
where $0<t<\infty$
and $x_\perp$ represents the coordinates of the 2d symmetric
plane. \footnote{This is a Bianchi I spacetime with an extra
  rotational symmetry in the $x_\perp$ plane. The authors of
  \cite{Gumrukcuoglu:2007bx} 
found two different solutions of this form for a universe with a pure cosmological constant.
By choosing $p_1=-\frac{1}{3}$ and $p_2=p_3=\frac{2}{3}$ in \eqref{kasner-ds}, 
the other planar branch of the Kasner-de Sitter solution
can be found in this gauge to be,
\begin{equation}
ds^2= -dt^2 +\left( {2\over 3} H^{-1} \cosh{{ 3 H  t}\over {2 }} \left(\sinh {{3 H t }\over {2 }}\right)^{-{1\over 3}}\right)^2dr^2 %dx^2
 + \left(\sinh {{3 H t}\over {2}}\right)^{4\over 3} %(dy^2 + dz^2) 
dx_\perp^2,
\end{equation} 
However, this spacetime is singular at $t=0$.}
This is indeed a very anisotropic spacetime near $t=0$ where the $r$ direction grows linearly with time while 
the expansion rate in the $x_\perp$ plane goes to zero, in other words, it becomes static. 

The authors in \cite{Gumrukcuoglu:2007bx} noted that $t=0$ is not a real singularity in this case, but just a coordinate singularity and the universe
near this point is represented by a non-singular Kasner solution of the form,
\begin{equation}
ds^2 \approx -dt^2 + t^2 dr^2%dx^2 + dy^2 + dz^2.
+dx_\perp^2.
\label{Milne2}
\end{equation}
This is in fact a piece of the 4d Minkkowski space.
Looking at the $t-r%x
$ subspace 
one identifies a 2d Milne space, which covers the interior of the future lightcone of any point
in the 2d Minkowski spacetime so one can perform a change of variables
that brings this metric into the usual 4d Minkowski space metric form. 
This means that the hypersurface of $t=0$ is not the real Big Bang singularity but
there is a part of spacetime that lies behind it.
One could be tempted to set the initial state of the vacuum at $t=0$
\cite{Gumrukcuoglu:2007bx,Pitrou:2008gk,Kim:2012bv,Dey:2013tfa,Dey:2012qp,Dey:2011mj},
but it is clear that in this geometry one should go beyond this hypersurface
since any disturbances can propagate freely through this null hypersurface all the way to us. 
We will show this in the following section.
 
\subsection{Maximally Extended Spacetime}

Taking into account the considerations made above regarding the $t=0$ hypersurface,
it is clear that one would like to know how to extend the spacetime beyond this point.
As shown in the Appendix B, there are several different gauges to describe the Kasner-de Sitter spacetime \eqref{Kasner-deSitter-1}.
In order to do this it will prove convenient to rewrite the metric given above in a different
gauge, namely,
\begin{equation}
ds^2= - {{dT^2}\over {f(T)}} + f(T) dr^2
%dx^2 
+ T^2  dx_\perp^2 % (dy^2 + dz^2) 
\label{Kasner-deSitter-2}
\end{equation}
where
\begin{equation}
\label{ft}
f(T): = H^2 T^2 - {{R_0}\over T}.
\end{equation}
In order for $T$ to be the timelike coordinate, and  $f(T)>0$, we take $L<T<\infty$, 
where
\bea%T_0 
\label{L}
L:= H^{-2/3} R_0^{1/3},
\eea
such that $f(%T_0
L)=0$.
In this form, the metric resembles the solution of a de Sitter black hole written in Schwarzschild coordinates, but 
there are several important differences. 
The first one is that the 
$x_\perp$ part of the metric represents a plane and it does not have a spherical 
symmetry as in the usual black hole geometries. This explains why there is no 
constant term in \eqref{ft}.  Furthermore this is a time-dependent solution so, as it is,
it only describes a region similar to the Schwarzschild-de Sitter (SdS) solutions beyond the
cosmological horizon. Finally, let us look at the relative sign of each of the terms on the function
$f(T)$. The first term describes the existence of a positive cosmological constant in our 
energy-momentum tensor as we should since we want our metric to approach de Sitter space asymptotically but the factor $R_0$ seems to
represents a negative mass term.\footnote{The case with positive mass term $R_0$ does not
have any horizon but only the singularity at $T=0$. This is in fact, the negative branch solution
mentioned earlier.} In this new coordinate system, the $t=0$ region is mapped into a finite time $T=L$,
\footnote{We can also set the value of $R_0$ to exactly match the solution found in
the previous form.} 
which in this language corresponds to the horizon of this geometry. In fact, it is the analog of the cosmological horizon in the SdS geometry. 

In summary, this metric can be thought of as describing the region behind a cosmological horizon of a planar black hole
of negative mass embedded in de Sitter space.
The detailed coordinate transformations between the two descriptions of the Kasner-de Sitter 
metric \eqref{Kasner-deSitter-1} and \eqref{Kasner-deSitter-2}
are explained in the Appendix B.

The important point of having this representation for our spacetime is that it becomes now clear
how to interpret the region {\it before} $t=0$. One should do what is normally done in the black hole
geometries to obtain the part of the spacetime beyond the horizon so that $T$ becomes spacelike and
$r$ becomes timelike in the region where $T<L$.  Replacing $T$ and $r$ with the new coordinates $R$ and $\tau$, respectively,
we can write the metric in the form,
\begin{equation}
ds^2=  - {\tilde f}(R) d\tau^2 
+ {{dR^2}\over {\tilde f(R)}} + R^2dx_\perp^2
\end{equation}
where $0<R<L$,
so that 
\begin{equation}
\tilde f(R) := H^2 \Big({L^3 \over R}-R^2\Big)>0.
\end{equation}
The other difference with the black hole case is that
$\tilde f(R)$ does not have another horizon, there is no other root of this function so the metric
plunges directly into a singularity at $R=0$.

The relevant question for us is to what extent this singularity could affect the 
initial conditions in our universe; the initial conditions for our anisotropic inflation. This is a question
about the causal structure of this spacetime which is better addressed in a Kruskal diagram. 
In order to do this we introduce the new coordinate system given by
\footnote{Look at the Appendix A
for the details of this coordinate transformation.},
\begin{figure}
\begin{centering}
\includegraphics[width=.8\linewidth,origin=tl]{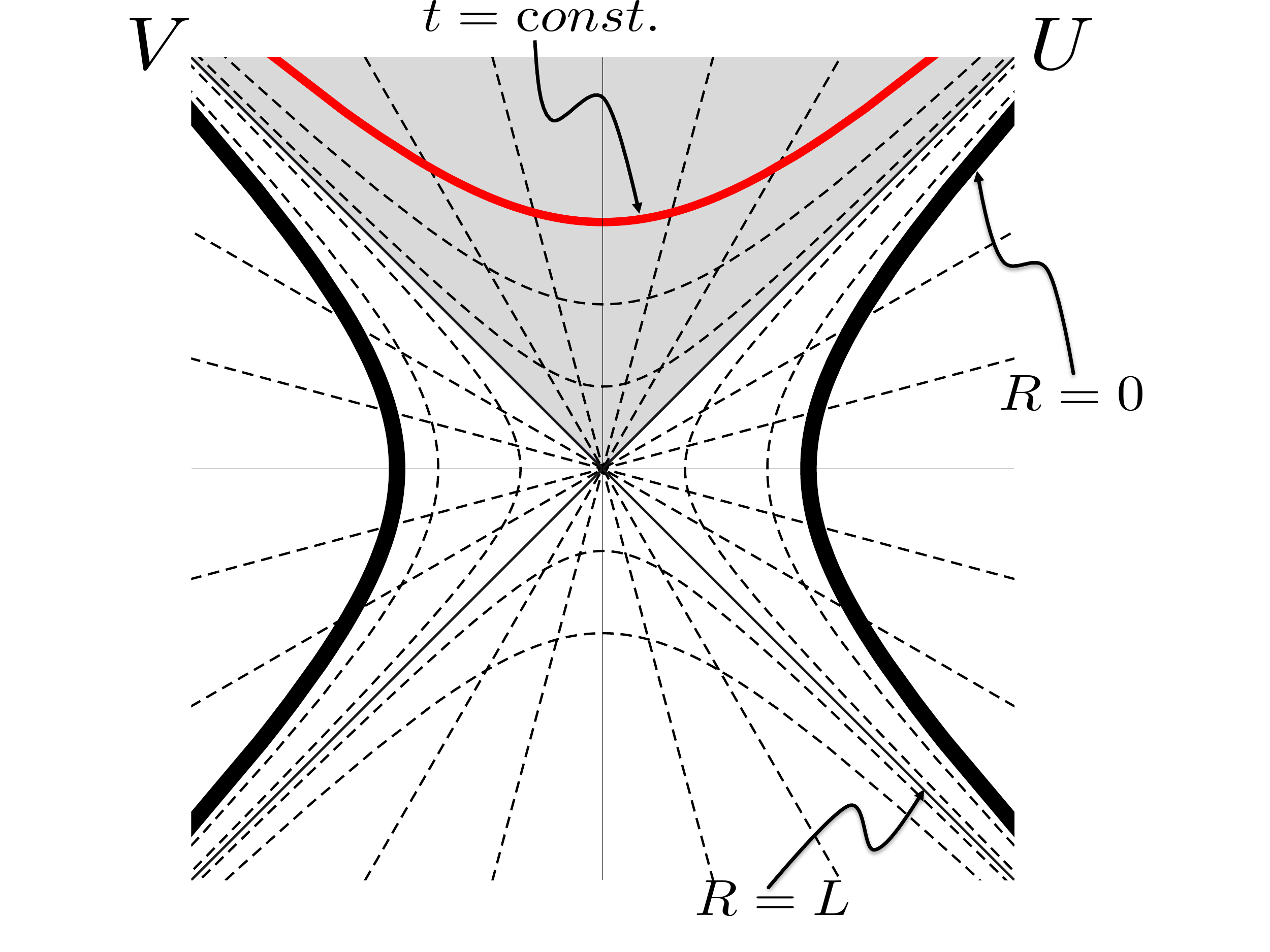}
\end{centering}
\caption{Kruskal diagram for the maximally extended Kasner-de Sitter solution. 
The shaded region is covered by the original Bianchi I  metric \eqref{Kasner-deSitter-1}
(and also by \eqref{Kasner-deSitter-2}). The lightlike lines at $R=L$ and the thick solid timelike curves at $R=0$
correspond to the cosmological horizon and the timelike singularity, respectively. The red curve denotes the constant
time hypersurface in the metric given by Eq. \eqref{Kasner-deSitter-1}. Each point in this diagram corresponds to a $2d$ plane.
}
\label{fig:KKdS2}
\end{figure}
\bea
\label{Kruskal}
ds^2=
-{\cal F}(R) dU dV + R^2dx_\perp^2,
\eea
where 
\bea
\label{F}
{\cal F}(R):=\left(\frac{2}{3 H L}\right)^2
{\rm exp}\left[
-\sqrt{3} {\rm arctan}
\left(\frac{L + 2 R} {\sqrt{3} L}\right) \right]
\frac{\big(R^2+ LR+ L^2\big)^{\frac{3}{2}}}{R},
\eea
and the coordinates $U$ and $V$ read
\bea
\label{UV_def}
V:={\rm exp}\Big(-{\frac{3 H^2 L}{2 }}\tau\Big) {\cal H}(%r
R)^{\frac{1}{2}},
\quad
U:=-{\rm exp}\Big({\frac{3 H^2 L}{2 }}\tau\Big) {\cal H}(%r
R)^{\frac{1}{2}},
\eea
with
\bea
{\cal H}(R) := {\rm exp}\left[\sqrt{3}{\rm arctan} 
\left(\frac{L+ 2R}{\sqrt{3}L}\right)\right]
\left(\frac {L-R}{\sqrt{R^2+ L R+L^2} }\right),
\label{hr}
\eea
and where $R$ is obtained from the relation,
\begin{equation}
U V = {\cal  H}(R).
\end{equation}

Using this coordinate system we see that nothing special happens at the horizon where $R=L$. On the other hand,
there is a real singularity that appears at $R=0$, a timelike singularity that lies on a hyperbolic line on the $U-V$ plane. 
Finally, hypersurfaces of constant $T$ (or constant $t$ in the original metric \eqref{Kasner-deSitter-2}) in the time-dependent part of the spacetime are given 
by hyperbolas in the interior of the lightcone.

Following a similar procedure as one does in
the Schwarzschild case one can find the maximal extension of this
geometry. We show in Fig. (\ref{fig:KKdS2}) the Kruskal diagram of 
this maximal extension. Its structure is similar to the Schwarzschild diagram rotated by 90 degrees.

We show in red a typical hypersurface of constant time in the cosmological part of the spacetime described by Eq. (\ref{Kasner-deSitter-1}). This can be thought of
a constant time hypersurface at the beginning of inflation. It is clear that the past lightcone of any point
in this hypersurface would intersect the timeline singularity (the thick black curves in the spacelike part of the
geometry) and therefore one cannot disregard its possible effect on the quantum state of the perturbations in our current universe.

\subsection{Quantum Initial State}

The study of cosmological perturbations in our background is complicated by the fact that
we are evolving not in an FRW universe but in an anisotropic Bianchi I universe. 
This brings the additional complication of the mixing of scalar- and tensor-type perturbations during the
initial anisotropic stage of the universe \cite{Gumrukcuoglu:2007bx,Pitrou:2008gk}.
In the following we will concentrate on the study of the perturbations of a massless scalar field and its
evolution on this background as a simplified model for perturbations. This is of course
an approximation and one should really perform the correct evolution of perturbations
along the lines of \cite{Gumrukcuoglu:2007bx,Pitrou:2008gk}. We leave this for future work.

The lesson drawn from the previous section is that to study the quantum state of the universe
in the cosmological region (the shaded region in Fig. (\ref{fig:KKdS2})), 
one should first understand the form of the metric in the spacelike 
region outside of the horizon. 
In order to do that let us start by writing the extension of the solution given by
Eq. (\ref{Kasner-deSitter-1}) beyond the horizon in a similar gauge, namely,
\begin{equation}
ds^2= - \left( {2\over 3} H^{-1} \sin{{ 3 H \rho}\over {2 }} 
\left(\cos {{3 H \rho }\over {2 }}\right)^{-{1\over 3}}\right)^2 d\tau^2 + d\rho^2+ \left(\cos {{3 H 
\rho}\over {2}}\right)^{4\over 3} dx_\perp^2.
\end{equation}
There are two relevant regions in this metric. 
The near horizon part of the geometry where $\rho\approx 0$ and the
metric approaches the Minkowski space in a Rindler form,
\begin{equation}
ds^2= - \rho^2 d\tau^2 + d\rho^2 + dx_\perp^2
\label{Rindler}
\end{equation}
and the singularity region at $%r=r
\rho=\rho_{max}:=\frac{\pi}{3H}$ where the metric approaches the Taub 
geometry \cite{Taub:1950ez}
\begin{equation}
ds^2= - (\rho_{max}-\rho)^{-2/3} d\tau^2 + d\rho^2 + (\rho_{max}-\rho)^{4/3}dx_\perp^2. 
\end{equation}

Not all timelike singularities are harmful and the quantization of a scalar field in this 
background could be possible if the information of the singularity were to be shielded
from the actual cosmological observers inside of the horizon. In order to investigate this possibility
we study the quantization of a massless scalar field in the vicinity of this Taub timelike singularity.
Following \cite{Blau:2006gd}, it would be useful to write this metric in the following gauge
$\xi:=(\rho_{\rm max}-\rho)^{4/3}$
\begin{equation}
ds^2= - \xi^{-1/2} \left(d\tau^2 - d\xi^2\right) + \xi %\left(dy^2 + dz^2 \right)
dx_\perp^2
\end{equation}
where the singularity occurs at $\xi \rightarrow 0$. 
To understand the behavior of the massless scalar field modes in this background, 
we start by decomposing the scalar field as
\begin{equation}
\phi_{k_\perp,E} = \xi^{-1/2} \psi_{k_\perp, E}(\xi) e^{-i k_{\perp} x_{\perp}} e^{-i E \tau}
\end{equation}
so the equation for the field $\psi_{k_\perp, E}$ takes the form,
\begin{equation}
\left[-{{d^2}\over {d\xi^2}}  + V_{k_\perp}(\xi) \right] \psi_{k_\perp,E}
 = E^2 \psi_{k_\perp,E}
\end{equation} 
where
\begin{equation}
\label{potential_singularity}
V_{k_\perp}(\xi) := - {{1} \over {4}} \xi^{-2} + k^2_{\perp} \xi^{-3/2}.
\end{equation}
This is a Schr\"{o}dinger type equation for the field $\psi$ with a divergent 
potential near $\xi=0$ where $V_{k_\perp}(\xi) \approx - (2 \xi)^{-2}$. Potentials
of this type have been discussed in the literature in \cite{Horowitz:1995gi,Blau:2006gd} where it was
argued that this potential would lead to $2$ normalizable solutions
near the singularity.
This means that we would have to
impose some sort of boundary condition at the singularity, making
the solution for the scalar field and, in turn, our quantum state
unpredictable.

Note that the coefficient in front of the $1/\xi^2$ term in the potential
is rather special since a slightly more negative value would lead to a much more serious
problem with an ill-defined quantum mechanical problem \cite{Landau:1990qp}.  We would of course
like to have a different type of singularity where the potential is not attractive but
repulsive so that the perturbations are uniquely specified by their asymptotic value
far away from the singularity. Taking into
account the backreaction of the scalar field in this
background may improve the behavior of the fluctuations near the singularity and 
achieve such a repulsive potential. This is precisely what one finds in another
closely related singular instanton, the so-called Hawking-Turok instanton \cite{Hawking:1998bn}. In this case one
can show that the situation improves dramatically taking into account backreaction. See the discussion
in \cite{Garriga:1998tm}. On the other hand, to study this in our case would require a careful treatment 
of the anisotropic nature of the scalar perturbations in this part of the geometry using a decomposition  
of the form described in \cite{Gumrukcuoglu:2007bx,Pitrou:2008gk}. We leave the investigation of this point for future work.

\section{Anisotropic Inflation as a result of Quantum Tunneling}

The arguments presented in the previous section show that one should take
into account the spacelike region of the geometry in order to describe the
quantum state of the cosmological fluctuations in the timelike region. On the other
hand, the extension of the simple Kasner-de Sitter geometry across 
the horizon yields a timelike singularity that is ``visible'' from our universe,
spoiling the predictability of this spacetime. One could think of different
ways to improve on this situation, either by introducing some regulating procedure,
by cutting out entirely the singularity from the spacetime \footnote{Similarly what is done for 
the Hawking-Turok instanton in \cite{Garriga:1998tm,Bousso:1998pk}.} or making it ``invisible'' by adding some other 
form of matter that dominates near the singularity \footnote{Another 
possibility would be to allow the spacetime to end similarly to what
happens in the ``bubble from nothing'' geometry \cite{BlancoPillado:2011me,Garriga:1998ri}.}.

In this paper we would like to take a different approach and give a new interpretation to the
anisotropic Kasner-de Sitter solution as the outcome of a quantum
tunneling process of a previously compactified space. In order
to make the connection to this interpretation we should first 
imagine that the coordinates $x_\perp$ 
are not infinite but compact such that collectively represent a $2d$ torus, $T_2$. 
This does not change anything in terms of the solutions presented 
earlier, they are still solutions of Einstein's equations with a 
pure cosmological constant. The difference is that we should
think of the spatial topology of the universe as $R \times T_2$
instead of $R^3$. This does not suppose a radical change at least
for the timelike part of the geometry where these directions are
expanding and this new view will only impose a 
minor restriction on the size of these extra dimensions over our
past cosmological history in order for them to be compatible
with phenomenology. Looking at the form of the solution near the $t=0$ we see that these
extra dimensions approach a static configuration. This suggests a possible modification of the region of the
space across the horizon that considers this spacelike region as a part of the universe 
where the extra-dimensions were static. These two regions together would therefore describe 
a decompactification transition.

These type of transdimensional transitions were first discussed in
models of flux compactification in \cite{BlancoPillado:2009di,Carroll:2009dn} in the context of a 
higher-dimensional landscape of multiple vacua. In \cite{BlancoPillado:2010uw,Graham:2010hh,Adamek:2010sg} the authors
discussed another example of decompactification from a lower
dimensional spacetime very similar to the one we have now.  
The difference between those 
models and the present work is the symmetry of the space. 
In their case, the spacetime had
anisotropic spatial curvature (it had either open or closed subspaces) that led
to a lower bound in the number of e-foldings after the anisotropic
initial expansion of the transition
\footnote{This is due to the limits on the induced quadrupole generated by the
late time anisotropic expansion due to curvature. See \cite{Demianski:2007fz,BlancoPillado:2010uw,Graham:2010hh,Adamek:2010sg} for details.}. This requirement made it difficult
to get an observational effect in the spectrum of cosmological perturbations since, as
we explained earlier, a large number of e-foldings would move all the effects from
primordial anisotropic inflation out of our present horizon.
Here on the other hand, we do not have this problem since both sections of spacetime
are spatially flat and the number of e-folds could be as low as 60.

The decompactification solutions presented in \cite{BlancoPillado:2010uw,Adamek:2010sg} were
described by instantons that also mediated the creation of black
objects in de Sitter space similar to the ones presented in \cite{BlancoPillado:2009mi,Carroll:2009dn}. In those cases the solutions had two different
horizons, a cosmological horizon that led to the anisotropic universe 
dominated by the cosmological constant and the ``black hole'' horizon
that would lead to a spacetime resembling a lower dimensional compactified universe.
In our case, we only have one horizon, the cosmological horizon. The difference
is again the lack of curvature in our spacetime, so we need to supply the solution
with some extra ingredient that allows for this other horizon that would
substitute and regularize the geometry in the spacelike region.
One can try to do this by adding an electric charge to this solution.
This can be accomplished 
easily in the {\it Schwarzschild-like} gauge by introducing a new term in the solution 
for $f(T)$ proportional to $Q^2/T^2$. We discuss this possibility in the Appendix C where we show
that this geometry also leads to a single horizon and an again a visible
singularity. The timelike continuation of this spacetime has been considered recently in
\cite{Chen:2013eaa,Chen:2013tna}. Much of our discussion in this paper applies to these solutions as well.
\begin{figure}[tbph]
\begin{center}
\includegraphics[width=0.7\linewidth,origin=tl]{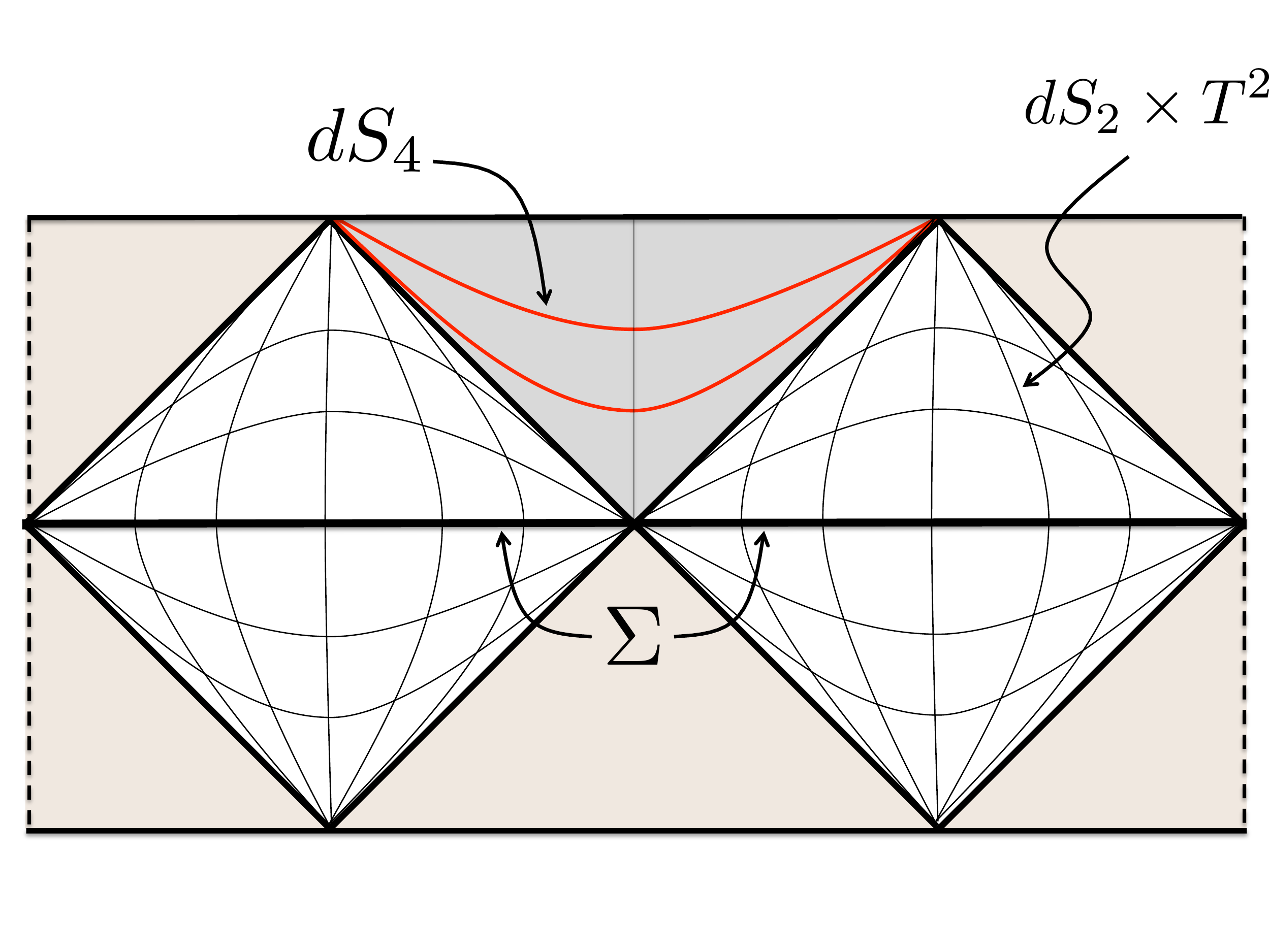}
\end{center}
\caption{The Penrose diagram for the Kasner-de Sitter bubble nucleation from a 
$\text{de~Sitter}_2 \times T_2$ parent vacuum. Each point in this figure corresponds 
to a $2d$ torus, the $T_2$. The region of the spacetime shaded in gray is the part 
of the metric described by Eq. (\ref{Kasner-deSitter-1}). The white region corresponds 
to compactified parent vacuum described by the metric in (\ref{dS2-metric-spacelike}). The asymptotic
constant time slices represented in red rapidly approach an isotropic de Sitter space in $4d$.} 
\label{ds2t2ds4}
\end{figure}

It is interesting to note that one could, in principle, obtain
an exact solution with the properties we are looking for by changing the sign
of the coefficient in front of the kinetic term for the Maxwell field. This is a rather 
exotic possibility and we will not consider it further in this paper.

A much more interesting possibility was discussed in \cite{Arnold:2011cz} where the
authors found the instanton transitions we are interested in considering
the contribution to the geometry from the Casimir type of calculation. It 
is temped to think of this geometry as the quantum corrected geometry
of the singular classical toroidal black hole in de Sitter space we have been 
discussing.

The Penrose diagram of this type of solution is shown in Fig. (\ref{ds2t2ds4}) where
we denote by $\Sigma$ the Cauchy surface for this geometry. In the 
following we will approximate the geometry in this spacelike region
by the simpler $dS_2 \times T_2$ solution. This corresponds, in fact,
to the Hawking-Moss limit of the instanton transition, where the 
size of the extra dimension does not change in the region between
the horizons. Written in this gauge the solution becomes,
\beq
ds^2= \left[- {{1}\over {H_{2d}^2}}\sin (H_{2d} \, r) ^2 d\tau^2 + dr^2 \right] + dx_\perp^2,
\label{dS2-metric-spacelike}
\eeq
which clearly takes the correct Rindler form given by Eq. (\ref{Rindler}) to
match to the cosmological Kasner-de Sitter solution across the lightcone.

\begin{figure}[tbph]
\begin{centering}
\includegraphics[width=0.7\linewidth,origin=tl]{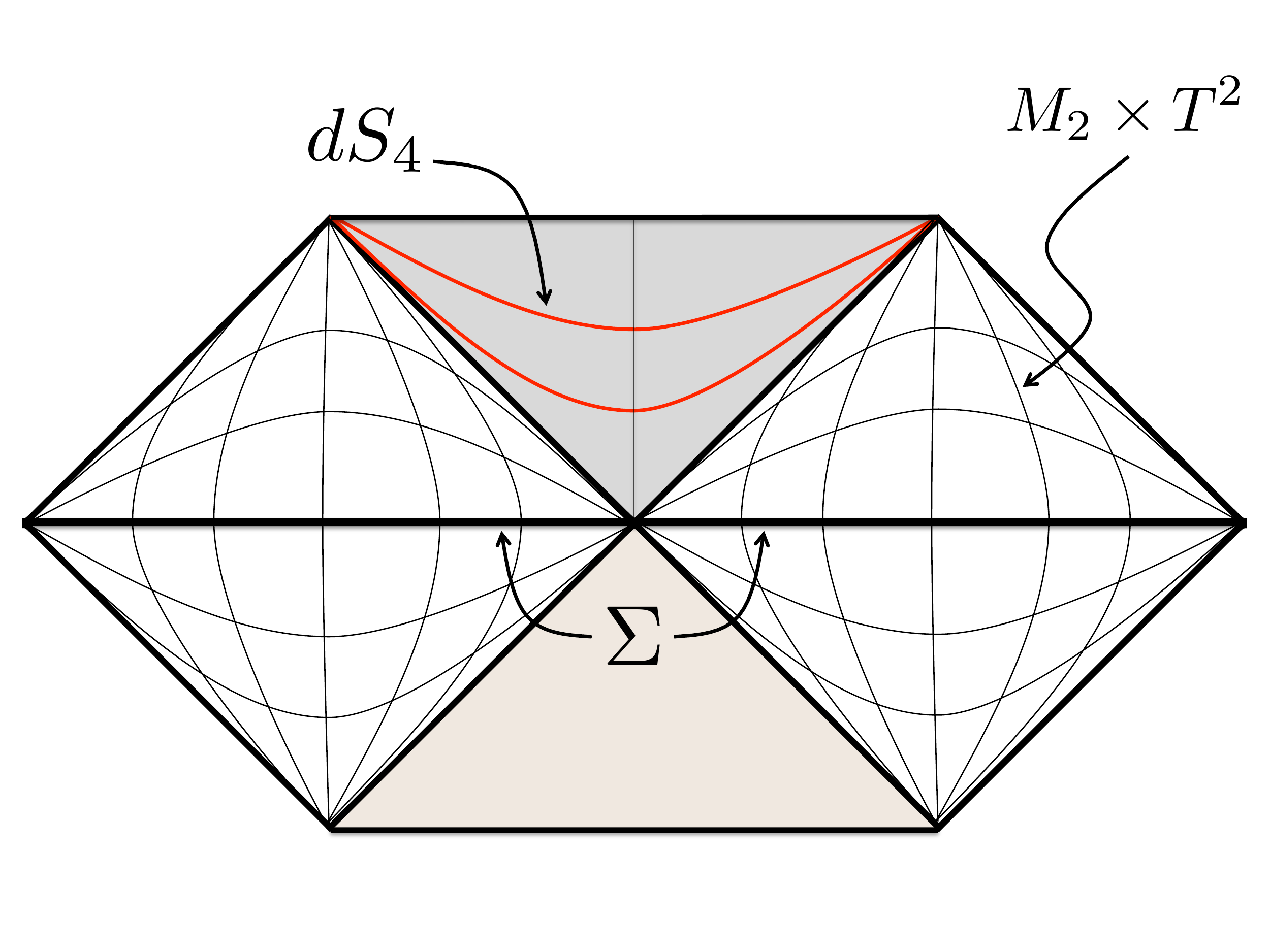}
\end{centering}
\caption{The Penrose diagram for the Kasner-de Sitter bubble nucleation from a $\text {Minkowski}_2 \times T_2$ parent vacuum. Each point
in this figure corresponds to a $2d$ torus, the $T_2$.} 
\label{m2t2ds4}
\end{figure}

Another possibility is to assume that the initial state of the
universe was in a static $M_2 \times T_2$ configuration right
before its decompactification transition. The metric outside of
horizon will now be given by  $M_2 \times T_2$ in Rindler coordinates,
\begin{equation}
ds^2= -r^2 d\tau^2 + dr^2 
+ dx_\perp^2 .
\end{equation}
This initial state is also compatible with the boundary conditions of the cosmological
evolution inside of our bubble and it is therefore worth considering even if its 
interpretation as a tunneling process is not so clear in this case. A diagram
of such a transition is given by Fig. (\ref{m2t2ds4}).

Here we do not specify the matter content that could give rise to 
these parent compactified states and simply assume that they exist. It is
also important to stress that in order to identify a model for this setup one would
also have to study its perturbative stability which in many simple models would not
be easy to achieve either for $dS_2 \times T_2$ or $M_2 \times T_2$.

Finally there is one more appealing point for this new interpretation of the non-singular Kasner-de Sitter solution. 
From the point of view of a generic Kasner solution 
the spacetime we discussed seems to be fine-tuned to avoid the initial 
Big-Bang singularity. On the other hand, the interpretation of the metric as a tunneling transition gives an
explanation for this rather unnatural initial conditions. The regularity 
of the instanton enforces the form of the solution around the 
lightcone and therefore its regularity is necessary in order
to be able to have a transition.

This new interpretation of the spacelike region of our solution
will allow us to set up the initial quantum state on the $\Sigma$ hypersurfaces.
This is what we do next.

\section{Quantization of a scalar field}

We are interested in understanding the spectrum of perturbations in this
background geometry. As a first approximation we will study the 
spectrum of fluctuations for a massless scalar field $\varphi$ 
minimally coupled to gravity: 
\beq
S_{\varphi} = \int{d^4x \sqrt{-g} \left(-{1\over 2} 
g^{\mu \nu} \partial_{\mu} \varphi \partial_{\nu} \varphi \right)}.
\eeq

Although we do not specify a concrete origin of this massless scalar field,
one can imagine that this is a simplified analog of the fluctuation of the inflaton field in this
anisotropic background \cite{Gumrukcuoglu:2007bx,Pitrou:2008gk}
or the isocurvature fluctuation of a subdominant light degree of freedom during inflation
which could be converted into the adiabatic perturbation after inflation \cite{BlancoPillado:2010uw}.

The background metric in the shaded region of Fig. (\ref{fig:KKdS2}) 
can be written in several different gauges as we described earlier. For
our purposes in this section it will be useful to express it in the following form\footnote{Look 
at the Appendix B for a detailed explanation of the change of coordinates among the 
different gauges used throughout this paper.}
\beq
ds^2 = -{{d\eta^2}\over {\sinh^2 (-H_{2d}\eta)}} + \alpha^4  
{{e^{4H_{2d}\eta/3}}\over {\sinh^{2/3} (-H_{2d} \eta)}}  dr^2 + \alpha^{-2}  {{e^{-2H_{2d}\eta/3}}\over {\sinh^{2/3} (-H_{2d} \eta)}} 
dx_\perp^2
\label{conformal-Kasner-deSitter}
\eeq
where $-\infty < \eta< 0$ and we have introduced the definitions, $\alpha=2^{1/3}$ and  $H_{2d} = 3% H_{4d}
H$. 

The expansion of the quantized scalar field in this region of spacetime can be given in general by,
\beq
\phi(\eta,r,x_{\perp}) = \int{dk \sum_{k_{\perp}} \left[{1\over {(2 \pi)^{3/2} }}  \tilde a_{k_{\perp},k} ~f_{k_{\perp},k}(\eta)  e^{i k_{\perp} x_{\perp}} e^{-i k r}  + \text{h.c}\right]}~,
\eeq
where the evolution equations of each mode functions are,
\begin{equation}
\label{evolution}
\left[{{d^2}\over {d\eta^2}}  + \Omega^2(k_{\perp},k,\eta) \right] 
f_{k_\perp,k}
(\eta) = 0
\end{equation} 
with,
\beq
\Omega^2(k_{\perp},k,\eta)  = \alpha ^{-4} \sinh ^{-4/3} (-H_{2d} \eta) ~~
e^{2 H_{2d} \eta /3} \left(\alpha^6 k_{\perp}^2 + e^{-2 H_{2d} \eta} k^2  \right)~.
\eeq
We have not been able to solve these equations analytically so we will integrate them numerically for each value of 
$k$ as well as $k_{\perp}$. This will allow us to compute the power spectrum of the perturbations outside of the
horizon. In order to do that we need to set up a vacuum state
or an initial form of the mode functions $f_{k_\perp,k}(\eta)$ in the limit of $\eta \rightarrow -\infty$.

\subsection{Choice of a parent vacuum}

Several groups have tackled the quantization of a scalar field in this geometry
using numerical \cite{Gumrukcuoglu:2007bx} 
as well as analytic techniques 
\cite{Kim:2012bv,Minamitsuji:2012ap,Kim:2011pt,Kim:2010wra,Dey:2013tfa,Dey:2012qp,Dey:2011mj}. The important difference
with our present work is the choice of the vacuum state which is now dictated by the parent vacuum
before decompactification.
Previous computations
followed the conventional approach of most inflationary models and assume that 
one could set the initial vacuum state by looking at the
form of the metric at $\eta \rightarrow -\infty$.
Taking the $\eta \rightarrow -\infty$ limit in Eq.  (\ref{conformal-Kasner-deSitter}) one arrives at,
\beq
ds^2 = {4\over  {e^{- 2 H_{2d}\eta }}}\left(- d\eta^2+ dr^2\right) +dx_\perp^2
\eeq
which is nothing more that our metric near the horizon given by Eq. (\ref{Milne2}) but written in 
a conformal gauge. In other words this is the conformal form of the spacetime given
by $\text{Milne}_2 \times T_2$. This is just a confirmation that indeed our metric becomes very anisotropic
as one goes far into the past as it is supposed to.

The use of $H_{2d}$ in this metric is arbitrary in this limit and can be reabsorbed in the definition of the  
coordinates. We use the $2d$ subscript to make the connection to the other possible parent vacua, 
namely the $dS_2 \times T_2$ case whose open slicing metric would take exactly the same form as 
the Milne case near the horizon, namely,
\beq
ds^2 = {{1}\over {\sinh^2 (-H_{2d} \eta )}} \left(- d\eta^2+ dr^2\right) +dx_\perp^2 
 \rightarrow {4\over  {e^{-2 H_{2d}\eta }}}\left(- d\eta^2+ dr^2\right) +dx_\perp^2. 
\eeq
This is again nothing surprising, we are just saying that an open universe slicing of $dS_2$ should
not know about the cosmological constant at early times, so it should behave 
as a spatially-flat universe dominated by curvature, a Milne universe in 2 dimensions in this case.

Let us discuss the two different vacua separately and identify the correct mode functions
for each case.

\subsubsection{$M_2 \times T_2$}

One can write the equations of motion for the perturbations near the lightcone to obtain
\begin{equation}
\left[{{d^2}\over {d\eta^2}}  + \left(4 e^{2H_{2d} \eta} k^2_{\perp} + k^2\right)\right]
f_{k_\perp,k} (\eta) 
= 0~,
\end{equation}
which can be thought of as the equations for a massive scalar field in $1+1$ Milne spacetime where
the mass scale is given by mode number along the internal dimensions, $k_{\perp} $.
The general solution of this equation can be written in terms of a combination of the 
Bessel functions of the form, $J_{\pm i \tilde k} \left(2 \tilde k_{\perp} e^{H_{2d}\eta}\right)$, where 
we have introduced the definitions, $\tilde k := {{ k}\over H_{2d}}$
and $\tilde k_{\perp} := {{ k_{\perp}}\over H_{2d}}$. 
Imposing that the mode functions behave as,
\begin{equation}
\lim_{\eta \rightarrow - \infty}
f_{k_\perp,k}(\eta) \propto {1\over {\sqrt{2k}}} e^{-ik \eta} 
\end{equation}
as one approaches $\eta \rightarrow -\infty$ (in other words $t\rightarrow 0$)  and normalizing them on any constant time slice
inside of the lightcone
\footnote{One can see by looking at Fig (\ref{m2t2ds4}) that these constant time slices are, in this case, Cauchy surfaces for the whole spacetime.}, 
one arrives at the following form for the mode functions,
\begin{equation}
f^{(c)}_{k_\perp,k}(\eta) = \sqrt{{{\pi}\over{2H_{2d} \sinh(\pi \tilde k)}} }J_{-i \tilde k} 
\left(2 \tilde k_{\perp} e^{H_{2d}\eta}\right)~,
\end{equation}
where the superscript $(c)$ refers to the fact that one can identify this choice of mode functions as the
usual conformal vacua of a Milne spacetime \cite{Birrell:1982ix}. The authors in
 \cite{Kim:2012bv,Minamitsuji:2012ap,Kim:2011pt,Kim:2010wra} found
that this vacuum leads to a divergent behavior in the limit of $k \ll k_{\perp}$, the 
so-called planar regime. Furthermore it has also been shown that this vacuum induces
a severe backreaction problem in this limit as well \cite{Dey:2011mj}. All these constraints
make it difficult to consider this vacuum state as the relevant one for our period of primordial
anisotropic inflation. 

It is also clear that one cannot consider this vacuum
as the one arising from a tunneling transition where the universe decompactifies from $M_2 \times T_2$
since things would blow up at the horizon \cite{Birrell:1982ix} preventing the existence of the instanton itself.
In fact, our interpretation of the anisotropic geometry in Eq. (\ref{Kasner-deSitter-1}) as the interior of a bubble created 
within a previously existing spacetime forces us to take the vacuum state of the parent vacuum as the correct state
for scalar perturbations. In our case, this is nothing more than the usual Minkowski two-dimensional vacuum written in this Milne coordinate system.
This corresponds to the mode functions of the form,
\beq
\label{minkvacuum}
f^{(M)}_{k_\perp,k}(\eta) 
= {1\over 2} \sqrt{{{\pi}\over {H_{2d}}}} e^{\pi \tilde {k}/2} 
~H_{i\tilde {k}}^{(2)} \left(2 \tilde {k}_{\perp} e^{H_{2d} \eta}\right)
\eeq
where $H_{i\tilde {k}}^{(2)}$ denotes the Hankel functions of the second kind. One can show that these 
mode functions are related to the previous ones by a Bogoliubov transformation of the form,
\begin{equation}
f^{(M)}_{k_\perp,k}
= \alpha _k f^{(c)}_{k_\perp,k}
 + \beta_k \left(f^{(c)}_{k_\perp,k} \right)^* 
\end{equation}
where
\beq
\alpha_{\tilde k} = {{e^{\pi {\tilde k}/2}}\over {\sqrt{e^{\pi \tilde k} - e^{-\pi \tilde k}}}} 
~~~~~ \beta_{\tilde k} =- {{e^{-\pi {\tilde k}/2}}\over {\sqrt{e^{\pi \tilde k} - e^{-\pi \tilde k}}}}~.
\eeq

We show in the Appendix D 
how one can obtain the explicit form of this vacuum state by studying the behavior 
of the mode functions on a Cauchy surface ($\Sigma$) on the previous vacuum and propagating them to the 
interior of the bubble.

\subsubsection{$dS_2 \times T_2$}

We will also consider the possibility that  our
parent vacuum was $dS_2 \times T_2$. Following techniques similar to the ones used in open inflation \cite{Garriga:1997wz,Garriga:1998he} one can show that
the correct vacuum state inside of our bubble is given by the analytic continuation
of the appropriate solutions in the spacelike region of the $dS_2 \times T_2$ geometry\footnote{This is $1+1$ dimensional
analogue of the Bunch-Davies vacuum in the open chart of de Sitter found in \cite{Sasaki:1994yt}.}.
We show the details of the calculation in the Appendix E. 
The resulting vacuum is given by,
\beq
\phi(\eta,r,x_{\perp}) = \int{dk \sum_{k_{\perp},i} \left[{1\over {(2 \pi)^{3/2} }}  \tilde a_{k_{\perp},k,i} ~f^{(i)}_{k_{\perp},k}(\eta)  e^{i k_{\perp} x_{\perp}} e^{-i k r}  + \text{h.c}\right]}~,
\eeq
where
\begin{eqnarray}
f^{(1)} _{k_{\perp},k}(\eta)&=& {1\over {\sqrt{2k}}}  {{e^{\pi k /2H_{2d}} }\over{\sqrt{2 \sinh (\pi k /H_{2d})}}} N(k,k_{\perp})~\tilde f^{(1)} _{k_{\perp},k}(\eta)\\
f^{(2)}_{k_{\perp},k}(\eta) &=&  {1\over {\sqrt{2k}}} {{e^{\pi k /2H_{2d}} }\over{
\sqrt{2 \sinh (\pi k /H_{2d})}}} \left( L(k,k_{\perp}) ~\tilde f^{(1)} _{k_{\perp},k}(\eta)+ e^{-\pi k/H_{2d} }
\tilde f^{(2)}_{k_{\perp},k}(\eta)\right)
\end{eqnarray}
and we have introduced the functions,
\begin{equation}
\tilde f^{(1)} _{k_{\perp},k}(\eta)= e^{-ik \eta} F\left[-\nu, \nu+1,1-\mu,{{1+\xi_i}\over 2}\right]
\end{equation}
\begin{equation}
\tilde f^{(2)} _{k_{\perp},k}(\eta)=  e^{ik \eta}F\left[-\nu, \nu+1,1+\mu,{{1+\xi_i}\over 2}\right]
\end{equation}
with
\begin{equation}
\xi_i = \coth (H_{2d} \eta)~~~;~~~
\mu = i\left( {k\over H_{2d}}\right)~~~;~~~
\nu (\nu +1) = - \left({{k_{\perp}}\over H_{2d}}\right)^2
\end{equation}
and $F$ denotes the generalized hypergeometric function so that $F[a,b,c,x] = {}_2F_1[a,b,c,x]$. Finally the normalization factors are given by,
\begin{eqnarray}
N(k,k_\perp) &=& {{\Gamma(1+\nu-\mu) \Gamma(-\mu-\nu)}\over {\Gamma(1-\mu) \Gamma(-\mu)}} \\
L(k,k_\perp) &=& - {{\Gamma(1+\mu) \Gamma(1+\nu-\mu) \Gamma(-\mu-\nu)}\over {\Gamma(1-\mu) \Gamma(-\nu) \Gamma(1+\nu)}}. 
\end{eqnarray}

\section{The Power Spectrum}

The results of the previous section give us the initial state of the scalar field modes right inside of the 
lightcone at the beginning of the bubble. One can then take these mode functions and evolve them forward in time numerically using
Eq. (\ref{evolution}). 

Here we present the results for the two cases we have studied, the $M_2 \times T_2$ and the
$dS_2 \times T_2$ parent vacua. For each case we give the power spectrum
\bea
{\cal P}
=
\frac{1}{2\pi^2}\big(\alpha^{-4}k^2 +\alpha^2 k_\perp^2\big)^{\frac{3}{2}}
\times
\begin{cases}
\big|f^{(M)} _{k_{\perp},k}(\eta\to 0)\big|^2
\quad\quad
(M_2\times T_2)    
\\
\sum_{i=1}^2\big|f^{(i)} _{k_{\perp},k}(\eta\to 0)\big|^2
\quad
(dS_2\times T_2)
  \end{cases}
\eea
 for several different
values of the angle $\theta$ as the function of $\bar k$,
which we define by the prescription, $ k = \bar k \cos \theta$ and
$k_{\perp} =\bar  k \sin \theta$ in the momentum space.

\begin{figure}[tbph]
\begin{center}
\begin{tabular}{ll}
\includegraphics[width=0.9\linewidth,origin=tl]{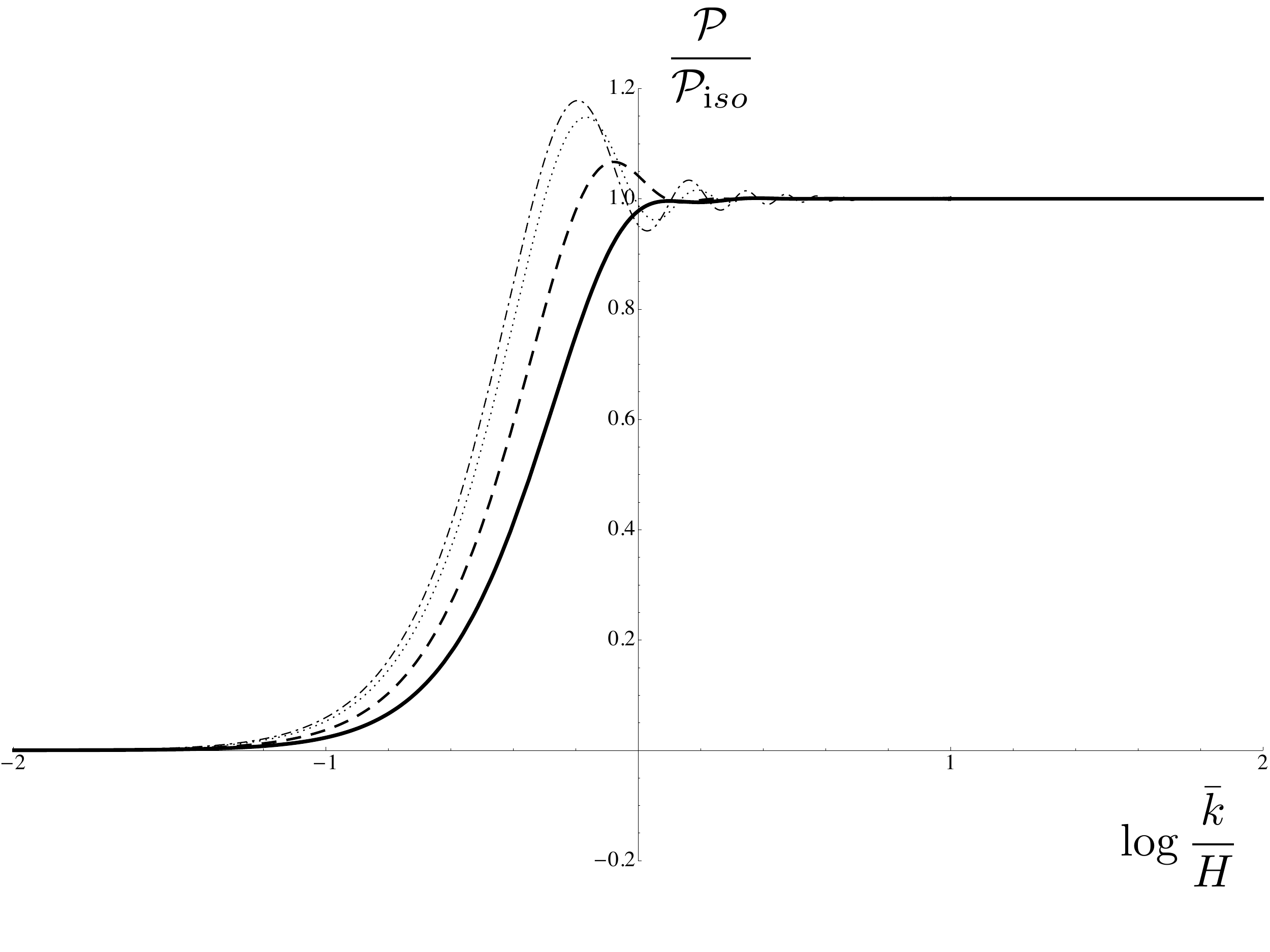}
\end{tabular}
\end{center}
\caption{Power Spectrum for the $M_2 \times T_2$. We plot the ratio of the power spectrum to the power 
in the isotropic limit as the function of $\bar k$. We plot the following angles,
$\theta={\pi/8,\pi/4,3\pi/8,\pi/2}$, which correspond to the solid, dashed, dotted and dot-dashed lines respectively.}
\end{figure}
We see that, as expected, the power spectrum becomes scale invariant and
isotropic when $\bar k$ becomes a few times larger than the corresponding comoving wavenumber associated with 
the comoving Hubble radius at the onset of the inflationary regime. These roughly correspond to the comoving momentum 
of the modes that leave the horizon when the universe starts to become isotropic and inflationary. To simplify our notation, 
in the following,  we normalize the comoving wavenumbers simply dividing them by $H$.

%after a short period of
%{\bf within a few periods of $\bar k$ after it becomes larger than the critical value}. 
This is basically due to the fact that the large
$\bar k$ modes leave the horizon when the universe is already expanding isotropically
so one should recover in this case the usual isotropic scale invariant spectrum of $dS_4$\footnote{In a realistic
model one should include a potential energy instead of a pure cosmological constant. That
would introduce to a small tilt in the scalar power spectrum.}.
Furthermore, the power in this case is not divergent as one approaches the
planar limit, $k \ll k_{\perp}$, as it is the case in the conformal vacuum.

In case of $M_2\times T^2$ parent vacuum,
it is interesting to note that the power spectrum is suppressed at large scales, low $\bar k$
and at the same time is slightly anisotropic in this regime, where the spectrum
is more sensitive to the initial anisotropic state as well as the anisotropic 
evolution. 

\begin{figure}[tbph]
\begin{center}
\begin{tabular}{ll}
\includegraphics[width=0.9\linewidth,origin=tl]{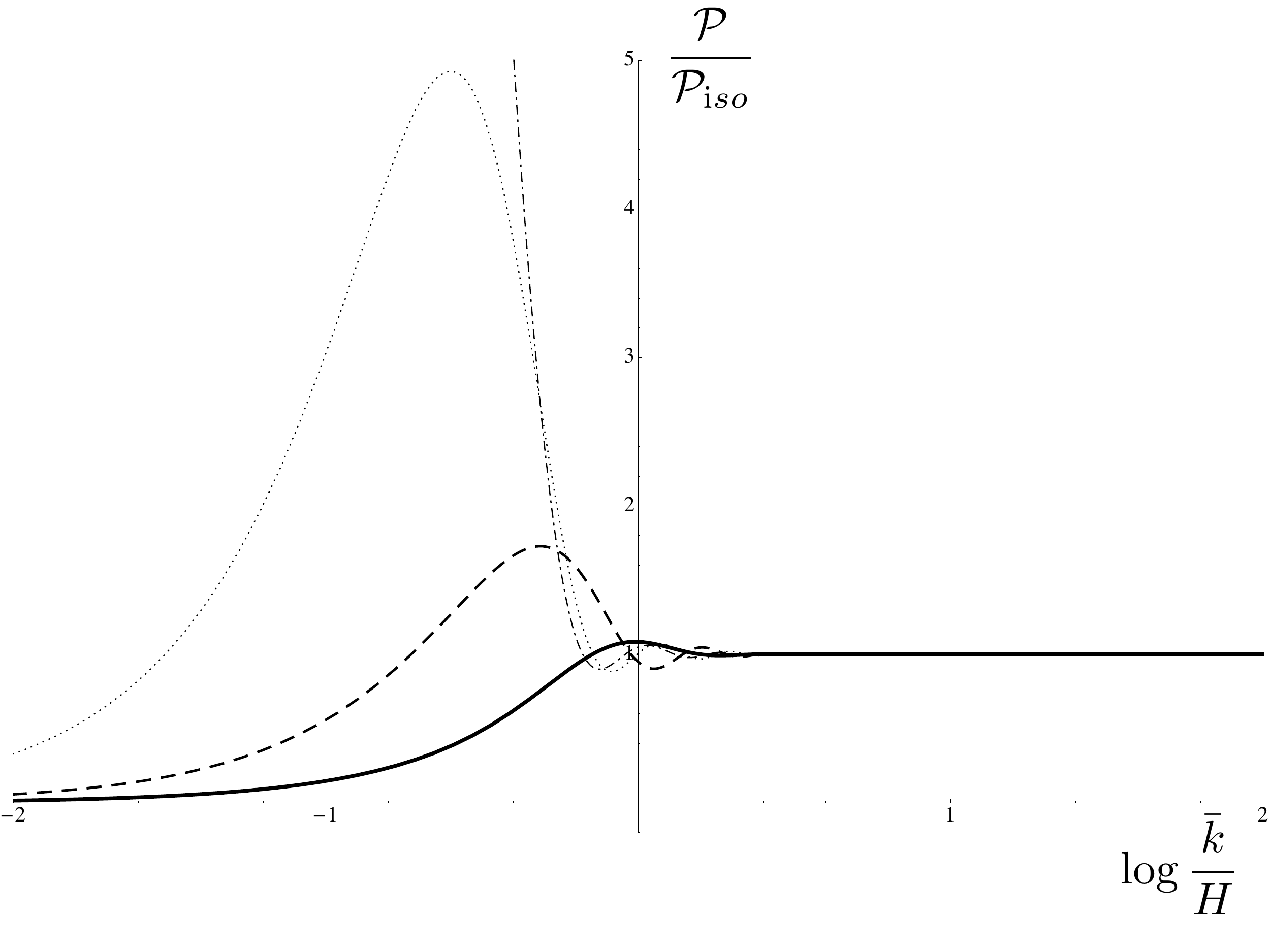}
\end{tabular}
\end{center}
\caption{Power Spectrum for the $dS_2 \times T_2$. We use the same set of angles as in the previous figure.} 
\end{figure}
The situation for the $dS_2 \times T_2$ parent vacuum is different and one finds a diverging
power in the planar regime for low $\bar k$. This does not signal the presence of any singularity behavior at the
lightcone, since the power spectrum is exactly the one that we will get for a set of massive
scalar fields in a pure $dS_2$ background. On the other hand, from the observational
point of view this type of spectrum with a large enhancement of power at low $\bar k$ seems
to be in contradiction with the CMB observations. This suggests that we should 
take the $\log\bar k\approx 0$ point in the figure to be associated with the largest observable scales
pushing all the extra power outside of our horizon today.
Unfortunately this also means that it would be very challenging to distinguish between this decompactification
model from any other model of nearly scale invariant isotropic spectrum in $4d$.

\section{Conclusions}

We have investigated the power spectrum of a massless scalar field in a model of
primordial anisotropic inflation given by the Kasner-de Sitter solution of Einstein's
equations in the presence of a pure cosmological constant. At early times,
this geometry approaches a state of high anisotropy where part of the metric
is described by a $1+1$ dimensional Milne universe while the other two spatial
directions remain static. Without any other sources present to preserve the anisotropy,
the spacetime quickly becomes the isotropic $dS_4$ making this metric a
good candidate for a primordial anisotropic stage in the early universe
before inflation. Any observable effect that one can obtain from this
initial period would therefore be encoded in the large scale today, the
scales that leave the horizon during the anisotropic era.
This means that these observables would be really sensitive to 
the initial vacuum state of the fluctuations. 

Previous attempts to study the fluctuations in this geometry rely on the
idea of identifying a vacuum state as the positive frequency modes for
the initial $\text{Milne}_2$ state. This led to several divergences on the
power spectrum that make this choice of initial state questionable. 

Here we present an alternative view on this Kasner-de Sitter geometry
that comes from the realization that the surface of the Big Bang for
this spacetime is in fact a coordinate singularity and nothing more than
the lightcone of the Milne slicing of Minkowski space. Extending the
geometry beyond this $t=0$ surface one encounters a timelike 
singularity that would be visible for observers in our cosmological 
spacetime making necessary to regulate the spacetime somehow
before we can identify a vacuum state for our perturbations.

We propose to give a different interpretation to the Kasner-de Sitter
metric as the outcome of a decompactification transition. We assume that
our parent vacuum state was described by a cosmological $1+1$ dimensional
spacetime
compactified over a $2d$ internal space that we take to be a flat torus, $T_2$.
For simplicity  we take the parent vacua to be either $\text{{\it de Sitter}}_2 \times T_2$ 
or ${\text {\it Minkowski}}_2 \times T_2$ such that they can be matched to the
Kasner-de Sitter metric along the lightcone. Taking into account the full geometry of the decay
process, one can identify a global Cauchy surface for these spacetimes and
obtain a suitable vacuum state for the scalar field perturbations. 

We calculated the power spectrum for a massless scalar field for different
orientations of the wavevector of the perturbation. We find that, as expected, 
the power spectrum is isotropic and scale invariant at small scales, 
since by the time that these modes leave the horizon the 
universe is pretty much isotropic. On the other hand, the spectra
vary substantially from these results for the large scales. We find that the spectrum
for $dS_2  \times T_2$ presents an important enhancement of its power
at large scales. This seems to be in contradiction with current CMB
observations that do not see this increase in power at large scales
but the opposite. One can of course assume that the number of
e-folds inside of our bubble was larger than $60$, which will
push the cosmological wavelengths associated with these 
modes outside of the current horizon making their effects 
almost irrelevant for us. 

We have also computed the power spectrum for a 
transition from $M_2  \times T_2$. The results in this
case are much more encouraging. We find that a transition 
of this kind leads to a suppression in the power spectrum 
at large scales as well a small variation on the power
with the angle, a small anisotropic effect. These 
features could be related to some of the low-$\ell$ anomalies
recently reported by the CMB collaborations.

In order to make a more precise comparison with the
data, and test the presence of detectable anisotropy in the
power spectrum one would have to compute the multiple correlators,
the $C_{\ell\ell' mm'}$, looking in particular for signals that could
set this model apart from other similar scenarios. Furthermore
one should also include the proper treatment of metric 
perturbations in these backgrounds using the results 
in \cite{Gumrukcuoglu:2007bx,Pitrou:2008gk} and study the
effect of considering new vacuum states coming from
decompactification. We leave these considerations for 
future research.\\

\section{Acknowledgements}
It is a pleasure to thank Jaume Garriga, Manuel Valle and Kepa Sousa for interesting discussions. 
We also
want to give special thanks to Mike Salem for collaboration in the early stages of this work and 
his invaluable guidance with the details of the calculation. J.J. B.-P. would also like
to acknowledge the hospitality of the Tufts Institute of Cosmology and the Yukawa Institute for
Theoretical Physics where part of this work was conducted. 
M. M. wishes to thank the University of the Basque Country for their hospitality. 
This work is supported in part by IKERBASQUE,
the Basque Foundation for Science and the Spanish Ministry of Science and Technology under Grant
FPA2012-34456 (J.J. B.-P),
the Yukawa Fellowship,
the Grant-in-Aid for Young Scientists (B) of JSPS Research under Contract No. 24740162
and the FCT-Portugal through Grant No. SFRH/BPD/88299/2012 (M.M.).

\appendix

\section{Kruskal extension of the Kasner-de Sitter universe}

Starting from the original metric
\bea
ds^2= 
-
H^2
\Big(\frac{%R_0
L^3}{R}-%H^2
R^2%r^2
\Big)d\tau^2
+ H^{-2}
 \Big(\frac{%R_0
L^3}{R}-%H^2r^2
R^2\Big)^{-1}dR^2%dr^2
+%r^2 
R^2dx_{\perp}^2,
\eea
where $L$ was defined in \eqref{L}, we can introduce the
following {\it tortoise coordinate} 
\bea
dR_\ast=\frac{dR}{H^2\big(\frac{L^3}{R}-R^2\big)},
\eea
to arrive to the metric of the form,
\bea
\label{toise}
ds^2=
H^2
\Big(\frac{%R_0
L^3}{%r
R}-%H^2
R^2%r^2
\Big)
\big(-d\tau^2+d%r
R_\ast^2\big)
+%r^2
R^2 dx_{\perp}^2,
\eea
where the relation between $R$ and $R\ast$ can be written explicitly as 
\bea
%r
R_\ast
=\frac{1} {3L H^2}
\left[
-\sqrt{3} {\rm arctan}
\left(\frac{1+\frac{ 2%r
R}{L}}{\sqrt{3}}\right)
+\ln\Big\{
\frac{\sqrt{R^2%r^2
+LR +L^2}}
      {L-R}%r}
\Big\}
\right]~. %,
\eea

\

The Kruskal extension can be obtained as follows:
Define new null coordinates
\bea
V=e^{k(\tau+%r
R_\ast)},\quad 
U= -e^{-k(\tau-%r
R_\ast)}\rightarrow
-dUdV=k^2 (-d\tau^2+d%r
R_\ast^2)e^{2k%r
R_\ast}.
\eea
Choosing $k$ so that 
$$
\frac{2k}{3L H^2}=-1,
$$
then
\bea
k^2e^{2k%r
R_\ast}
=\Big(\frac{3H^2 L %R_0^{\frac{1}{3}} H_i^{\frac{4}{3}} 
}{2}\Big)^2
{\rm exp}\left[
\sqrt{3} {\rm arctan}
\left(\frac{1+\frac{2R%r
}{L}}
{\sqrt{3}}\right)
\right]
\frac{L-R}%r}
{\sqrt{R^2%r^2
+L R%r
+L^2  }}.
\eea
which brings the metric \eqref{toise} to its final form, \eqref{Kruskal}-\eqref{hr}.
\bea
ds^2=
-{\cal F}(%r
R) dU dV+%r^2 (dy^2 +dz^2)
 R^2dx_\perp^2,
\eea
where 
\bea
{\cal F}(%r
R):=
% H^{-2} 
\left(\frac{2}{3 H%^2
L}
\right)^2
{\rm exp}\left[
-\sqrt{3} {\rm arctan}
\left(\frac{L + 2 %r
R}
{\sqrt{3} L}
\right)
\right]
\frac{\big(R^2+ LR+ L^2\big)^{\frac{3}{2}}}{R},
\eea
and the coordinates $U$ and $V$ satisfy,
\begin{equation}
U V =
{\cal  H}(R).
\end{equation}
with
\bea
{\cal H}(R) :=
{\rm exp}\left[\sqrt{3}{\rm arctan} 
\left(\frac{L+ 2R}%r}
{\sqrt{3}L}\right)\right]
\left(\frac {L-%r
R}{\sqrt{%r
R^2+ L%r
R+L^2} }\right),
\eea

\section{The Kasner-de Sitter metric in several gauges}

As we discussed in the main part of the text, perhaps the simplest version of the
Kasner-de Sitter solution can be written in a Schwarzschild-like metric of the
form,
\begin{equation}
ds^2= - {{dT^2}\over {f(T)}} + f(T) d\tilde r^2 
+ T^2 d\tilde x_\perp^2 
\end{equation}
where,
\begin{equation}
f(T) = H^2 \Big(T^2 - {%{R_0}
L^3 \over T}\Big),
\end{equation}
and $L$ was introduced in \eqref{L}.

We can now go to a proper time coordinate by defining,
\begin{equation}
t = {2\over 3} H^{-1}  \log \left[ \left({T\over L}\right)^{3/2} + \sqrt{\left({T\over L}\right)^3 -1}\right]
\end{equation}
to arrive at the metric,
\begin{equation}
ds^2= -dt^2 + (HL)^2
\left( \sinh{{ 3 H t}\over {2 }} \left(\cosh {{3 H t }\over {2 }}\right)^{-{1\over 3}}\right)^2 d\tilde r^2%x^2 
+ L^2 \left(\cosh {{3 H t}\over {2}}\right)^{4\over 3} d\tilde x_\perp^2~.
\end{equation}
Rescaling the spacelike coordinates we arrive to the desired form of the metric, given in the main text, namely
\begin{equation}
ds^2= -dt^2 + \left({2\over 3} H^{-1}  \sinh{{ 3 H t}\over {2 }} \left(\cosh {{3 H t }\over {2 }}\right)^{-{1\over 3}}\right)^2 %dx^2
dr^2 + \left(\cosh {{3 H t}\over {2}}\right)^{4\over 3}dx_\perp^2~.
\end{equation}

Finally we can also introduce another form of the metric by changing the time coordinate according to the identification,
\begin{equation}
\sinh \left(3 H t\right) = {1\over {\sinh (-3 H\eta)}}
\end{equation}
which brings the metric to the form,
\beq
ds^2 = -{{d\eta^2}\over {\sinh^2 (-3H\eta)}} + \alpha^4  {{e^{4H \eta}}\over {\sinh^{2/3} (-3 H \eta)}}  dr^2 + \alpha^{-2}  {{e^{-2H\eta}}\over {\sinh^{2/3} (-3 H \eta)}} %(dy^2 + dz^2)
dx_\perp^2.
\eeq

\section{Adding an electric field}

The form of our anisotropic metric in its Schwarzschild gauge suggests an
extension of this family of solutions that includes the possibility of an electromagnetic
field, in other words, there may be solutions of the following action
\bea
\label{em}
S&=&\int d^4x\sqrt{-g}
\left[
\frac{1}{2}R
-\frac{1}{4}F_{\mu\nu}F^{\mu\nu}
-\Lambda\right]
\eea
that respect our Bianchi I form and that have some sort of electromagnetic field turned on in its background.
Similarly to what one does in the case of black
holes with spherical horizon we take the ansatz
\beq
ds^2= - f(T)^{-1} dT^2 + f(T) %dx^2
dr^2 + T^2 dx_\perp^2%(dy^2 + dz^2)
\eeq
where the solution we are looking for looks like,
\beq
f(T) = H^2 T^2 -{{R_0}\over {T}}- {{Q^2}\over{T^2}}~.
\eeq 
This is indeed a solution of Einstein's equations for a configuration with a 
cosmological constant, $\Lambda = 3H^2$ and an electric field along the 
$r$ direction, namely
\beq
F_{Tr} =\frac{Q}{T^2}.
\eeq

The solutions of this family have many similarities to the ones presented in the main part of the
text and lead to the same kind of observational signatures. It is easy to see that this is indeed the 
case if we take the particular case where $R_0=0$ and bring the metric on
a more cosmological form, namely,
\beq
ds^2 = -dt^2 + a^2(t) dr^2%dx^2 
+ b^2(t) dx_\perp^2%(dy^2 + dz^2)
\eeq
where in this case we have the scale factors,
\bea
a(t)&=&2^{\frac{1}{2}}
             e^{H t_b}
\sinh\big(2H (t-t_b)\big)
\big[\cosh\big(2H (t-t_b)\big)\big]^{-\frac{1}{2}},
\quad
b(t) = 2^{\frac{1}{2}}
             e^{H t_b} \big[\cosh (2H (t-t_b))\big]^{\frac{1}{2}}. \nonumber
\eea
where $t_b = {1\over {2H}} \log \left({Q\over {2H}}\right)$.
We can bring the metric to a form much closer to the Kasner-de Sitter metric used in the
text by shifting the time as well as rescaling the coordinates to arrive to:
\begin{equation}
ds^2= -dt^2 + \left({1\over {2 H}} \sinh (2 H t) \left[\cosh (2 H t )\right]^{-{1\over 2}}\right)^2 %dx^2
dr^2 + \cosh (2 H t) dx_\perp^2~, %(dy^2 + dz^2)~,
\label{Kasner-deSitter-11}
\end{equation}
which clearly has the same qualitative behavior as one approaches the
lightcone, at $t=0$, as our Kasner-de Sitter geometry. In this regard, this metric
is another example of a primordial stage of anisotropic inflation. 

Note that this is the same metric as the one that has recently been discussed in 
the context of anisotropic inflation in \cite{Chen:2013eaa,Chen:2013tna} where the form of the
scale factors in this case were given in a slightly different form, 
\bea
a(t)&=&A(t)\frac{1-\frac{Q^2}{4A(t)^4 H^2}} 
        {\sqrt{1+\frac{Q^2}{4A(t)^4 H^2}}},
\quad
b(t)=A(t)\sqrt{1+\frac{Q^2}{4A(t)^4 H^2}},
\eea
with $A(t) = e^{H t}$.
One can investigate the
Kruskal diagram for this metric following exactly the same steps as we
have done in this paper and see that the structure for this metric is identical
to the Kasner-de Sitter case (See Fig. (\ref{fig:KKdS2})). 
This also implies that one should look beyond
the lightcone to set the initial quantum state of the perturbations. 

Studying the perturbations of a scalar field near the singularity one can check that 
the coefficient of the $\xi^{-2}$ term in \eqref{potential_singularity}
is now replaced with $-\frac{2}{9}$.
This again admits the existence of 2 normalizable solutions
near the singularity which implies that
this is a visible singularity for observers in our anisotropic universe.

We can include both effects in the metric, namely a mass and
an electric field but the conclusions would be unchanged so it
does not seem possible to regularize this singularity adding just
an electromagnetic field. One should therefore think about a consistent way to
specify the vacuum state in these kinds of models \cite{Chen:2013eaa,Chen:2013tna}
as well.

\section{The $M_2 \times T_2$ parent vacuum}

\subsection{Rindler and Milne spaces}

The full Minkowski space in $1+1$ dimensions can be split into 4 regions, 2 Rindler wedges and 2 Milne wedges. 
In this section we present the metric appropriate for each region, their relation to the usual Minkowski metric
and the analytic continuations that allow us to go between the different wedges.
\begin{figure}[tbph]
\begin{center}
\begin{tabular}{ll}
\includegraphics[width=0.76\linewidth,origin=tl]{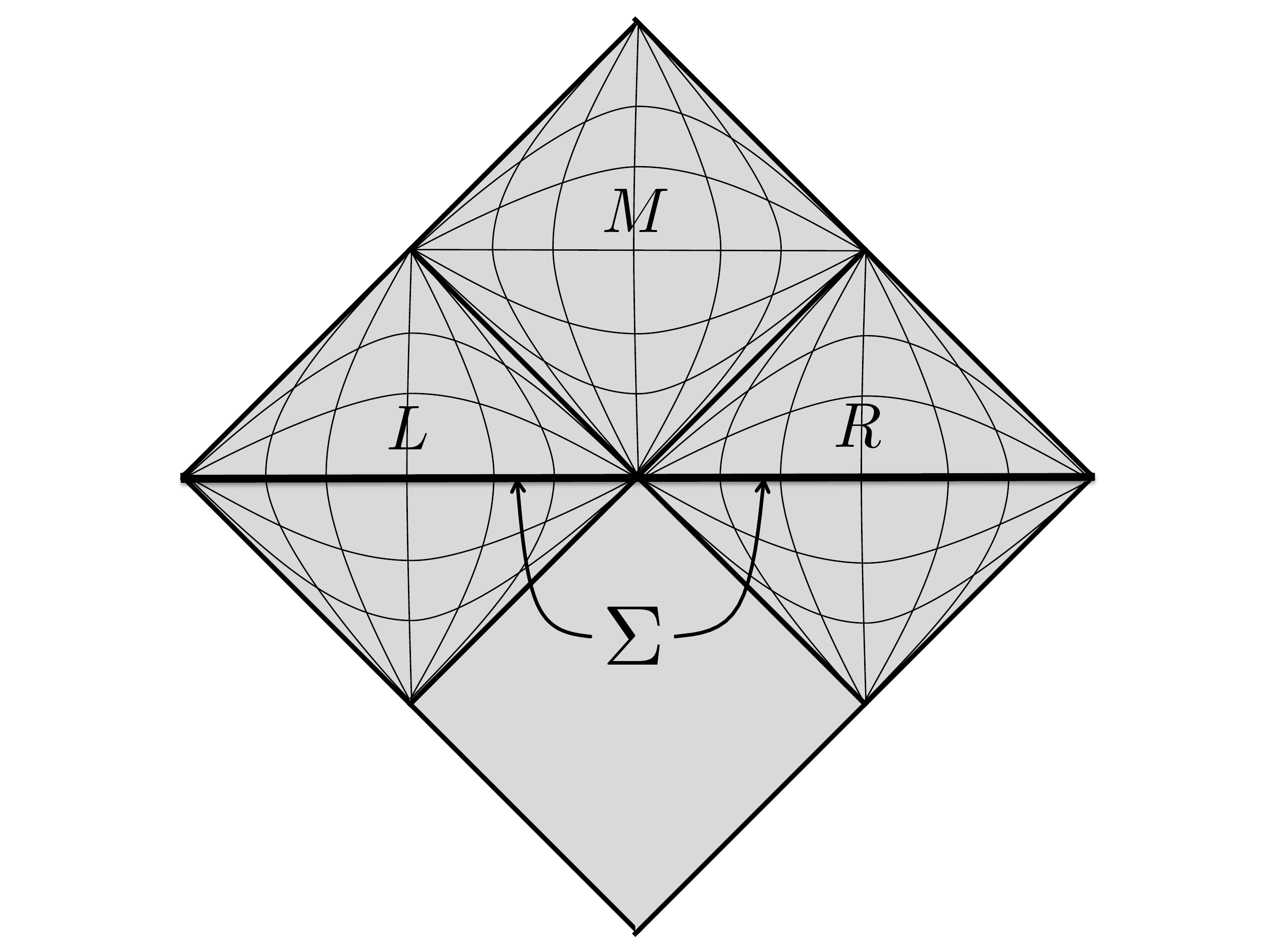}
\end{tabular}
\end{center}
\caption{The full Minkowski space in $1+1$ dimensions
split into 4 regions, 2 Rindler ($R$ and $L$) wedges and 
2 Milne wedges (the future wedge is $M$). We also show the Cauchy surface $\Sigma$ where
we will quantize the field.} 
\label{rindler_milne}
\end{figure}

The two Rindler wedges can be found by the embedding in Minkowski space by the following expressions,
\begin{eqnarray}
T = {%1
2\over H_{2d}}~e^{H_{2d}\eta_R} \sinh( H_{2d}  %\chi
r_R),\quad
X = {%1
2\over H_{2d}}~e^{H_{2d}\eta_R} \cosh( H_{2d}  %\chi
r_R)
\end{eqnarray}
where $-\infty < \eta_R < \infty$ and $-\infty < %\chi
r_R < \infty$.
\begin{eqnarray}
T = -{%1
2\over H_{2d}}~e^{H_{2d}\eta_L} \sinh (H_{2d} %\chi
r_L),\quad
X = - {%1
2\over H_{2d}}~e^{H_{2d}\eta_L} \cosh( H_{2d}  %\chi
r_L)
\end{eqnarray}
where $-\infty < \eta_L < \infty$ and $-\infty < %\chi
r_L < \infty$.

Furthermore the Milne part can also be found by the expression,
\begin{eqnarray}
T = {%1
2\over H_{2d}}~e^{H_{2d}\eta_M} \cosh(H_{2d} %\chi
r_M),\quad
X = {%1
2\over H_{2d}}~e^{H_{2d}\eta_M} \sinh (H_{2d} %\chi
r_M)
\end{eqnarray}
where $-\infty < \eta_M < \infty$ and $-\infty < %\chi
r_M < \infty$.

We can go from the R region to the M region by the following analytic continuation,
\begin{eqnarray}
\eta_R = \eta_M + {{\pi}\over {2H_{2d}}} i, \quad
%\chi
r_R = %\chi
r_M - {{\pi}\over {2H_{2d}}} i,
\end{eqnarray}
similarly for the L region we have,
\begin{eqnarray}
\eta_L = \eta_M + {{\pi}\over {2H_{2d}}} i, \quad
%\chi
r_L = %\chi
r_M + {{\pi}\over {2H_{2d}}} i.
\end{eqnarray}

\subsection{Scalar field quantization in $M_2 \times T_2$ }
The form of the metric for the $M_2 \times T_2$ parent vacuum state can be written for the region outside of the lightcone as
the so-called Rindler wedge,
\beq
ds^2=4
 e^{2H_{2d}\eta_R} \left(d\eta_R^2 -dr_R^2\right) + dx_\perp^2~.
\eeq
This metric however only covers half of the spacetime outside of the bubble, and one 
needs to supplement it with the analogous {\it left} wedge (L). We will see shortly the relevance
of these 2 different parts of the spacetime, but for the time being let us discuss the 
quantization of a massless scalar field on the  {\it right} wedge.

Using the same expansion for the scalar field as before,
\beq
\phi(\eta_R,r_R,x_{\perp}) = \int{dk \sum_{k_{\perp}} \left[{1\over {(2 \pi)^{3/2} }} 
 \tilde c_{k_{\perp},k} ~h_{k_{\perp},k}(\eta_R)  e^{i k_{\perp} x_{\perp}} e^{-i k r_R}  + \text{h.c}\right]}~,
\eeq
we get,
\beq
h_{k_\perp,k}'' + \left(k^2 - 4
e^{2H_{2d} \eta_R} k_{\perp}^2 \right) h_{k_\perp,k} = 0.
\eeq
One can think of this equation as the Schr\"{o}dinger equation for an exponential potential and an
energy state denoted by $k^2$. The solutions of these mode functions are therefore suppressed
for $\eta \rightarrow \infty$. One can show that the solutions that fulfill this requirement
are given by,
\beq
h_{k_\perp,k}^{(R)}(\eta_R) = \sqrt{{{2\sinh (\pi k /H_{2d})}}\over {\pi H_{2d}}}~ 
K_{ik/H_{2d}} \left({2{k_{\perp} e^{H_{2d} \eta_R}}\over H_{2d}}\right),
\eeq
where we have fixed the normalization so that the functions are normalized on the Rindler wedge. Putting this
together with the rest of the spatial dependence one arrives at,
\beq
\phi^{(R)}(\eta_R,r_R,k,k_{\perp})= \sqrt{{{2\sinh (\pi k /H_{2d})}}\over {\pi H_{2d}}}~ K_{ik/H_{2d}} 
\left({2{k_{\perp} e^{H_{2d} \eta_R}}\over H_{2d}}\right) ~e^{-ik r_R} .
\eeq
One can also define the corresponding functions on the left wedge by making the substitution $R \rightarrow L$
with $e^{-ik r_R}\to e^{ikr_L}$, 
as in the left wedge
time runs in the opposite way and so the positive frequency mode functions have the opposite sign in the exponent. 
Each of these functions only has support on their respective wedge but one can define a new mode function
defined over the entire Cauchy surface, $\Sigma$, that one obtains by merging the $r_R=0$ and
$r_L=0$ hypersurfaces (See Fig (\ref{rindler_milne})). 
These new functions are given by \cite{Unruh:1976db},
\beq
h^{(M)}_{k_\perp,k} =
  {1\over{\sqrt{2 \sinh (\pi k /H_{2d})}}} \left[e^{\pi k /2H_{2d}} \phi^{R}(\eta_R,r_R,k,k_{\perp}) 
+ e^{-\pi k /2H_{2d}}  (\phi^{L}(\eta_L,r_L,k,k_{\perp}))^*\right].
\eeq

Where the overall factor has been chosen so that the new modes are properly normalized on $\Sigma$. Furthermore
due to the nice analytic properties of these new functions one can analytically continue them into the 
interior of the lightcone to obtain the final form of our mode functions, namely,
\beq
f^{(M)}_{k_\perp,k}
(\eta_M) = {1\over 2} \sqrt{{{\pi}\over {H_{2d}}}} e^{\pi k/2H_{2d}} 
~H_{ik/H_{2d}}^{(2)} \left({{2k_{\perp} e^{H_{2d} \eta_%B
M}}\over H_{2d}}\right)~.
\eeq

These are in fact the mode functions that describe the usual Minkowski vacuum in the Milne coordinates so the 
expression for the initial state near the lightcone should be given by,
\beq
\phi(\eta_M,r_M,x_{\perp}) = \int{dk \sum_{k_{\perp}} \left[{1\over {(2 \pi)^{3/2} }}  \tilde a_{k_{\perp},k} ~f^{(M)}_{k_{\perp},k}(\eta_M)  e^{i k_{\perp} x_{\perp}} e^{-i k r_M}  + \text{h.c}\right]}~.
\eeq

\section{The $dS_2 \times T_2$ parent vacuum}
\subsection{$dS_2$ metrics.}

The Euclidean description of $dS_2$ is given by a 2 sphere which can be embedded in $3d$ by the equation
$X_0^2 + X_1^2 + X_2^2 = H_{2d}^{-2}$, which we can parametrize by,
\begin{eqnarray}
X_0 = H_{2d}^{-1} \cos \tau \cos \rho,\quad
X_1 = H_{2d}^{-1}  \sin \tau, \quad
X_2 =  H_{2d}^{-1}  \cos \tau \sin \rho,
\end{eqnarray}
giving the metric
\begin{equation}
ds^2_E =H_{2d}^{-2}  \left(d\tau^2 + \cos^2 \tau ~ d\rho^2\right).
\end{equation}

\begin{figure}[tbph]
\begin{center}
\begin{tabular}{ll}
\includegraphics[width=0.7\linewidth,origin=tl]{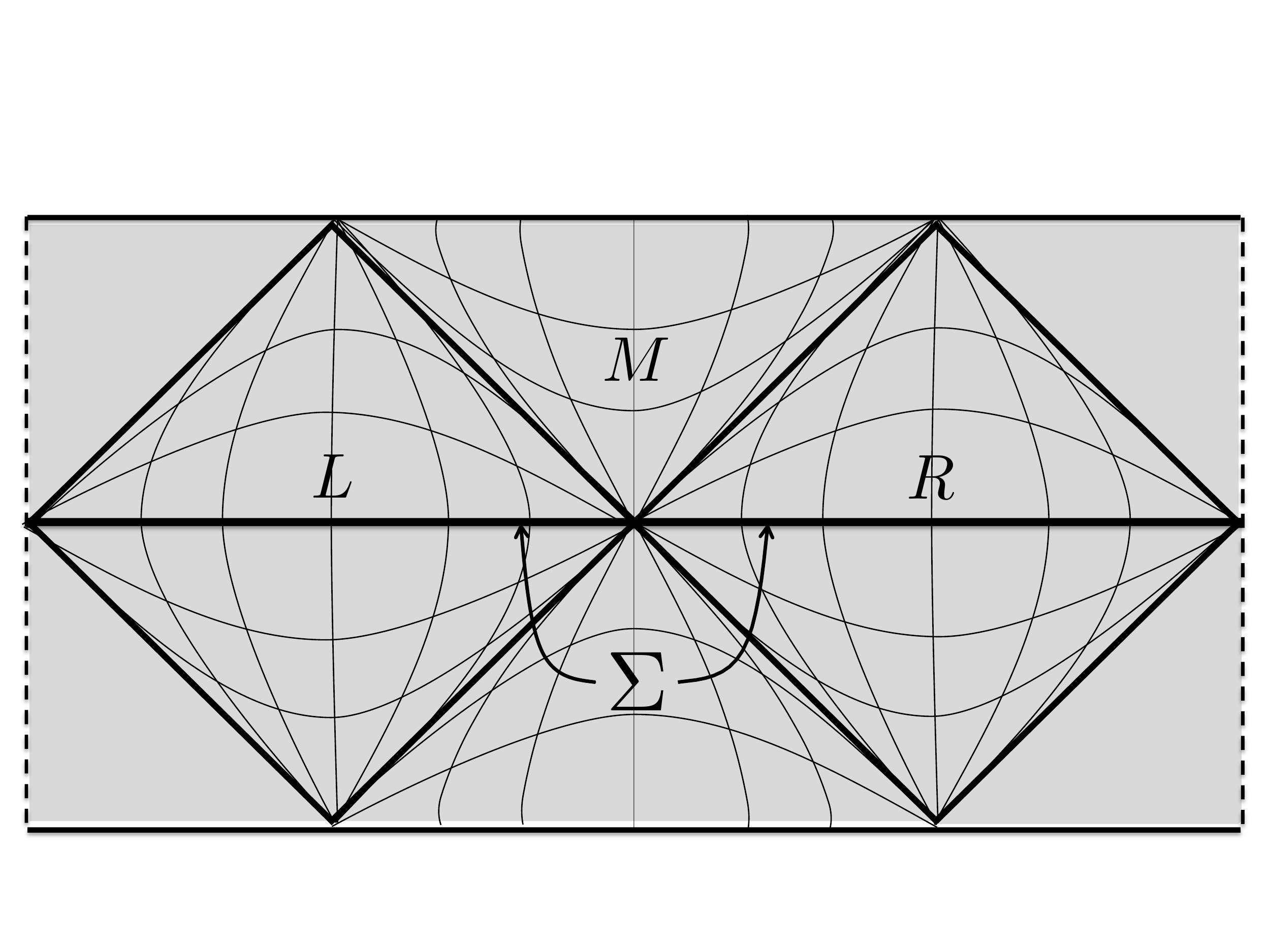}
\end{tabular}
\end{center}
\caption{The Penrose diagram for the full de Sitter space in $1+1$ dimensions split into the different regions discussed in the text.
} 
\label{Penrose-dS2}
\end{figure}

We can do the following Wick rotation,
\begin{eqnarray}
\tau = -i H_{2d} t_M + {{\pi}\over 2}, \quad
\rho = - i H_{2d} r_M
\end{eqnarray}
to obtain the embedding of part of $dS_2$ in $3d$ Minkowski space given by,
\begin{eqnarray}
X_0 &=& H_{2d}^{-1} \cosh (H_{2d} r_M) \sinh (H_{2d} t_M),\quad
X_1 = H_{2d}^{-1} \cosh (H_{2d} t_M), 
\nonumber\\
X_2 &=& H_{2d}^{-1}  \sinh (H_{2d} t_M) \sinh(H_{2d} r_M),
\end{eqnarray}
so the induced metric becomes,
\begin{equation}
ds^2_M =- dt_M^2 + \sinh^2(H_{2d} t_M) ~ dr_M^2.
\end{equation}
This slicing of $dS_2$ covers the region that we denote by $M$ in Fig. (\ref{Penrose-dS2}) \footnote{The label of this
region as $\text{M}$ comes from the fact that indeed this metric approaches the Milne one in the limit of  $t_M \rightarrow 0$.}

On the other hand, we can also do another Wick rotation,
\begin{eqnarray}
\tau = H_{2d} t_R,\quad
\rho = - i H_{2d} r_R + {{\pi}\over 2}
\end{eqnarray}
to obtain,
\begin{eqnarray}
X_0 &=& H_{2d}^{-1} \cos (H_{2d} t_R) \sinh ( H_{2d} r_R),
\quad
X_1 = H_{2d}^{-1} \sin (H_{2d} t_R),
\nonumber
 \\
X_2 &=& H_{2d}^{-1}  \cos(H_{2d} t_R) \cosh (H_{2d} r_R),
\end{eqnarray}
that gives the metric
\begin{equation}
ds^2_R = dt_R^2 -  \cos^2 (H_{2d} t_R) ~ dr_R^2.
\end{equation}
which covers the region $R$ in Fig. (\ref{Penrose-dS2}). We can also find the analogue of
this coordinate system for the $L$ part as well.

One can change from one chart to the other by doing,
\begin{eqnarray}
t_R = -i t_M + {{\pi}\over {2 H_{2d}}}, \quad
r_R = r_M -  i {{\pi}\over {2 H_{2d}}}~.
\end{eqnarray}

Also, we can introduce the a new coordinate system, by the following change of variables
\beq
\sinh (H_{2d} t_M) = {1\over {\sinh (-H_{2d} \eta_M)}}
\eeq
with $0 \le t_R \le \infty$ being mapped to $-\infty \le \eta_R \le 0$, to get,
\begin{equation}
ds^2_M = {1\over {\sinh^2 (-H_{2d} \eta_M)}} (-d\eta_M^2 +  dr_M^2)
\end{equation}
which is the metric induced by the following embedding,
\begin{eqnarray}
X_0 &=& H_{2d}^{-1} \cosh (H_{2d} r_M) ~\text{csch} (-H_{2d} \eta_M)\\
X_1 &=& H_{2d}^{-1} \cosh (H_{2d} t_M) \\
X_2 &=&  H_{2d}^{-1}  \sinh (H_{2d} t_M)~\text{csch} (-H_{2d} \eta_M)~.
\end{eqnarray}

Similarly, one can change coordinates outside of the horizon using the
following relation,
\beq
\cos (H_{2d} t_R) = {1\over {\cosh (H_{2d} \eta_R)}}
\eeq
with $-{{\pi}\over 2H_{2d}} \le t_R \le {{\pi}\over 2H_{2d}}$ 
being mapped to $-\infty \le \eta_R \le \infty$, so
we get,
\begin{equation}
ds^2_R = {1\over {\cosh^2 (H_{2d} \eta_R)}} (d\eta_R^2 -  dr_R^2).
\end{equation}
This form of the metric is given by the embedding,
\begin{eqnarray}
X_0 &=& H_{2d}^{-1} ~\text{sech} (H_{2d} \eta_R) \sinh ( H_{2d} r_R)\\
X_1 &=& H_{2d}^{-1}  \tanh (H_{2d} \eta_R)  \\
X_2 &=& H_{2d}^{-1}   ~\text{sech} (H_{2d} \eta_R) \cosh (H_{2d} r_R).
\end{eqnarray}

We can relate both charts by the following analytic continuation,
\begin{eqnarray}
r_R =r_M -  i {{\pi}\over {2 H_{2d}}}, \quad
\eta_R = -\eta_M -  i {{\pi}\over {2H_{2d}}}.
\end{eqnarray}

\subsection{Scalar Field Quantization in $dS_2 \times T_2$}

\subsubsection{Outside of the horizon}

As we described in the main part of the text, we will define our vacuum state
on the Cauchy surface $\Sigma$ outside of the horizon of the bubble that describes the 
tunneling process (See Fig. (\ref{ds2t2ds4})).  Our assumption is that
the spacetime in this region is well approximated by the $R$ and $L$ patches of $dS_2 \times T^2$ where one
can identify the analogous Cauchy surface. (See Fig. (\ref{Penrose-dS2}).)
Following the description in the previous Appendix E. 1. one can write the metric
of this part of $dS_2$ in the following way,
\begin{equation}
ds^2 = {1\over{\cosh^2(H_{2d} \eta_R)}} (-dr_R^2 + d\eta_R^2) +dx_\perp^2.% (dy^2+dz^2)
\end{equation}

In order to describe the vacuum state for a massless scalar field in this
geometry, we expand the field in the following form,
\beq
\phi(\eta_R,r_R,x_{\perp}) = \int{dk \sum_{k_{\perp},i} 
\left[{1\over {(2 \pi)^{3/2} }}  \tilde c_{k_{\perp},k,i} ~h_{k_{\perp},k}^{(i)} (\eta_R)  
e^{i k_{\perp} x_{\perp}} e^{-i k r_R}  + \text{h.c}\right]}~,
\eeq
so the equations of motion for the mode functions become
\footnote{Note that this is analogous to the
Schr\"{o}dinger equation for a $k^2_{\perp}/\cosh^2(x)$ potential.}, 
\begin{equation}
\left[- {{d^2}\over{d\eta_R^2}} + \left({{k^2_{\perp}}\over {\cosh^2(H_{2d} \eta_R)}} - k^2\right)\right] h_{k_\perp,k}^{(i)}(\eta_R)=0~.
\end{equation}
We can write two independent solutions of this equation in terms of Hypergeometric functions 
of the form,
\begin{equation}
\tilde h^{(1)}_{k_\perp,k} = \left({{\xi_o+1}\over {1-\xi_o}}\right)^{\mu/2} F\left[-\nu, \nu+1,1-\mu,{{1-\xi_o}\over 2}\right]
\end{equation}
\begin{equation}
\tilde h^{(2)}_{k_\perp,k}
= \left({{\xi_o+1}\over {1-\xi_o}}\right)^{-\mu/2} F\left[-\nu, \nu+1,1+\mu,{{1-\xi_o}\over 2}\right]
\end{equation}
where we have defined,
\begin{equation}
\xi_o= \tanh(H_{2d} \eta_R),
\quad
\mu = i\left( {k\over H_{2d}}\right),
\quad
\nu (\nu +1) = - \left({{k_{\perp}}\over H_{2d}}\right)^2~,
\end{equation}
and we have simplified the notation by defining the generalized hypergeometric function simply by $F[a,b,c,x] = {}_2F_1[a,b,c,x]$.
One now needs to find the correct combination of these functions
that are Klein-Gordon normalized on our Cauchy surface. As a first 
step in this direction, we identify the asymptotic form for each of the
mode functions.

In the $\eta_R \rightarrow \infty$ limit, the $\xi_o\rightarrow1$ one 
simply finds,
\begin{equation}
\tilde h^{(1)}_{k_\perp,k}
\rightarrow \left({{\xi_o+1}\over {1-\xi_o}}\right)^{\mu/2} = e^{ik \eta_R},
\quad
\tilde h^{(2)}_{k_\perp,k} 
\rightarrow \left({{\xi_o+1}\over {1-\xi_o}}\right)^{-\mu/2} = e^{-i k \eta_R}.
\end{equation}

To calculate the $\xi_o\rightarrow-1$ limit we use the relation of the hypergeometric functions,
\begin{eqnarray}
F[a,b,c,x] &=& {{\Gamma(c) \Gamma(c-a-b)}\over {\Gamma(c-a) \Gamma(c-b)}} F[a,b,a+b-c+1,1-x] 
\nonumber \\
&+& (1-x)^{c-a-b} {{\Gamma(c) \Gamma(a+b-c)}\over {\Gamma(a) \Gamma(b)}} F[c-a,c-b,c-a-b+1,1-x] .
\end{eqnarray}
so the mode functions in the  $\eta_R \rightarrow -\infty$  ($\xi_o\rightarrow -1$)
limit become,
\begin{eqnarray}
\tilde h^{(1)}_{k_\perp,k} &\rightarrow &\left({{\xi_o+1}\over {1-\xi_o}}\right)^{\mu/2} \left[{{\Gamma(1-\mu) \Gamma(-\mu)}\over {\Gamma(1+\nu-\mu) \Gamma(-\mu-\nu)}}  + {{\Gamma(1-\mu) \Gamma(\mu)}\over {\Gamma(-\nu) \Gamma(1+\nu)}} \left({{1+\xi_o}\over 2}\right)^{-\mu} \right] 
\nonumber \\
 &\approx &    {{\Gamma(1-\mu) \Gamma(-\mu)}\over {\Gamma(1+\nu-\mu) \Gamma(-\mu-\nu)}}~e^{i k\eta_R}
 + {{\Gamma(1-\mu) \Gamma(\mu)}\over {\Gamma(-\nu) \Gamma(1+\nu)}} ~e^{- i k \eta_R}
\end{eqnarray}
as well as,
\begin{eqnarray}
\tilde h^{(2)}_{k_\perp,k} &\rightarrow& \left({{\xi_o+1}\over {1-\xi_o}}\right)^{-\mu/2} \left[{{\Gamma(\mu) \Gamma(1+\mu)}\over {\Gamma(1+\nu+\mu) \Gamma(\mu-\nu)}}  + {{\Gamma(1+\mu) \Gamma(-\mu)}\over {\Gamma(-\nu) \Gamma(1+\nu)}} \left({{1+\xi_o}\over 2}\right)^{\mu} \right]
\nonumber \\
 &\approx & {{\Gamma(\mu) \Gamma(1+\mu)}\over {\Gamma(1+\nu+\mu) \Gamma(\mu-\nu)}} ~e^{- i k \eta_R} 
 + {{\Gamma(1+\mu) \Gamma(-\mu)}\over {\Gamma(-\nu) \Gamma(1+\nu)}} ~e^{i k \eta_R}.
\end{eqnarray}

Looking at these asymptotic expansions one can then identify the correct combination of the 
mode functions that are normalized in other words that satisfy the conditions,
\beq
\int_{-\infty}^{\infty}{d\eta ~h^{(i)}_{k_{\perp},k}  \left(h^{(i')}_{k'_{\perp},k'}\right)^*}  = 2\pi ~\delta  (k - k')
\delta_{i,i'}~.
%\delta(i - i')~.
\eeq
Using techniques borrowed from the $1d$ quantum mechanics problems \cite{Garriga:1998he,Garriga:1997wz} one can find the correct
combination to be\footnote{In particular, one can find this solution for the $1/\cosh^2(x)$ potential in \cite{Landau:1990qp}.},
\begin{eqnarray}
h^{(1)}_{k_\perp,k} = N(k,k_{\perp})~\tilde h^{(1)}_{k_\perp,k}, \quad 
h^{(2)}_{k_\perp,k} = L(k,k_{\perp})~\tilde h^{(1)}_{k_\perp,k} + \tilde h^{(2)}_{k_\perp,k}
\label{modefunctions}
\end{eqnarray}
where we have introduced the coefficients,
\begin{eqnarray}
N (k,k_{\perp})
&=& {{\Gamma(1+\nu-\mu) \Gamma(-\mu-\nu)}\over {\Gamma(1-\mu) \Gamma(-\mu)}} \\
L (k,k_{\perp})
&=& - {{\Gamma(1+\mu) \Gamma(1+\nu-\mu) \Gamma(-\mu-\nu)}\over {\Gamma(1-\mu) \Gamma(-\nu) \Gamma(1+\nu)}} .
\end{eqnarray}

\iffalse
One can of course verify that these combinations have the correct asymptotics

\begin{eqnarray}
h^{(1)}_{k_\perp,k} &\rightarrow&   {{\Gamma(1+\nu-\mu) \Gamma(-\mu-\nu)}\over {\Gamma(1-\mu) \Gamma(-\mu)}}  ~e^{i w \eta_C} ~~~~ (\eta_C \rightarrow \infty)\\\\
h^{(1)}_{k_\perp,k} &\rightarrow&  e^{i w \eta} + {{\Gamma(1+\nu-\mu) \Gamma(-\mu-\nu) \Gamma[\mu]}\over {\Gamma(-\nu) \Gamma(1+\nu) \Gamma[-\mu]}} ~~ e^{-i w \eta_C}  ~~~~ (\eta_C \rightarrow - \infty)\\
\end{eqnarray}
and
\begin{eqnarray}
h^{(2)}_{k_\perp,k} &\rightarrow&   -{{\Gamma(1+\mu) \Gamma(1+\nu-\mu) \Gamma(-\mu-\nu)}\over {\Gamma(1-\mu) \Gamma(-\nu) \Gamma(1+\nu)}}  ~e^{i w \eta_C} + e^{-i w \eta_C} ~~~~ (\eta_C \rightarrow \infty) \\\\
h^{(2)}_{k_\perp,k} &\rightarrow&  {{\Gamma(1+\nu-\mu) \Gamma(-\mu-\nu) }\over  {\Gamma(1-\mu) \Gamma(-\mu)}} ~~ e^{-i w \eta_C}  ~~~~ (\eta_C \rightarrow - \infty)\\
\end{eqnarray}
\fi

\subsubsection{Normalizing the mode functions}

The general expression for the mode decomposition in the R-region,  outside of the bubble, is given by,
\begin{eqnarray}
\phi(\eta_R,r_R,x_{\perp}) &=&  \int{dk \sum_{k_{\perp},i} \left[\tilde c_{k_{\perp},k,i} ~\phi^{(i)}_{k_{\perp},k}(\eta_R,r_R,x_{\perp}) + \text{h.c}\right]} 
\nonumber \\
&=&
\int{dk \sum_{k_{\perp},i} \left[\tilde c_{k_{\perp},k,i} ~{\cal N}_{k_{\perp},k}^{(i)}
~h^{(i)}_{k_{\perp},k}(\eta_R)  
 e^{-i k r_R}  e^{i k_{\perp} x_{\perp}}+ \text{h.c}\right]}
\end{eqnarray}
in order to quantize this model we need to normalize these modes using the Klein-Gordon normalization
given by,
\beq
\left(\phi^{(i)}_{k_{\perp},k}, \phi^{(i')}_{k'_{\perp},k'}\right) = -i \int{d\Sigma_{\mu} g^{\mu \nu} \left( \phi^{(i)}_{k_{\perp},k} \partial_{\nu} \left(\phi^{(i')}_{k'_{\perp},k'}\right)^* - \partial_{\nu}\phi^{(i)}_{k_{\perp},k} \left(\phi^{(i')}_{k'_{\perp},k'}\right)^*\right)}.
\eeq

Taking the Cauchy surface as the hypersurface of nucleation $\Sigma$ with normal vector in the R-region
\beq
n^{\mu} = \cosh(H_{2d} \eta_R)~(1,0,0,0)
\eeq
we find that $d\Sigma^{\mu} =  \cosh(H_{2d} \eta_R) \delta^{\mu}_0 ~d\Sigma = d^3 x \delta^{\mu}_0  $. Inserting this into the normalization expression, we get
\begin{eqnarray}
\left(\phi^{(i)}_{k_{\perp},k}, \phi^{(i')}_{k'_{\perp},k'}\right) &=& -i \int{d^3 x \left( \phi^{(i)}_{k_{\perp},k} \partial_{r_R} \left(\phi^{(i')}_{k'_{\perp},k'}\right)^* - \partial_{r_R} \phi^{(i)}_{k_{\perp},k} \left(\phi^{(i')}_{k'_{\perp},k'}\right)^*\right)}
\nonumber \\
 &=& -i ~{\cal N}_{k_{\perp},k}^{(i)}
           ~{\cal N}_{k'_{\perp},k'}^{(i')\ast}
 \left[2 i k ~ \left(\int{d^2 x_{\perp} e^{i (k_{\perp} - k'_{\perp})x_{\perp}} }\right)~ \left( \int{d\eta_R ~h^{(i)}_{k_{\perp},k}  \left(h^{(i')}_{k'_{\perp},k'}\right)^*}\right)\right]
\nonumber \\ 
 &=& 2 k \left(2 \pi\right)^3 
\Big|{\cal N}_{k_{\perp},k}^{(i)}\Big|^2 \delta (k_{\perp} - k'_{\perp}) \delta(k - k')% \delta(i - i')
\delta_{i , i'},
\end{eqnarray}
where to get to the last line, we use,
\beq
\int{d^2 x_{\perp} e^{i (k_{\perp} - k'_{\perp})x_{\perp}} } = (2\pi)^2 ~\delta (k'_{\perp} - k_{\perp}) 
\eeq
as well as,
\beq
\int_{-\infty}^{\infty}{d\eta_C ~h^{(i)}_{k_{\perp},k}  \left(h^{(i')}_{k'_{\perp},k'}\right)^*}  = 2\pi ~\delta  (k - k') \delta_{i , i'}.
\eeq

This means that we should take 
${\cal N}_{k_{\perp},k}^{(i)} =   \left({(2 \pi)^{3/2} \sqrt{2k}}\right)^{-1} $, so that our normalized functions become,
\beq
\phi^{(i)}_{k_{\perp},k} = {1\over {(2 \pi)^{3/2} \sqrt{2k}}}  h^{(i)}_{k_{\perp},k}(\eta_R)  e^{-i k r_R}
e^{i k_{\perp} x_{\perp}} ~.
\eeq
This calculation shows that the final expansion of the quantized field in the R region should be of the form,
\beq
\phi(\eta_R,r_R,x_{\perp}) =  \int{dk \sum_{k_{\perp},i} \left[{1\over {(2 \pi)^{3/2} \sqrt{2k}}}  \tilde c_{k_{\perp},k,i} ~h^{(i)}_{k_{\perp},k}(\eta_R) e^{-i k r_R} e^{i k_{\perp} x_{\perp}}  + \text{h.c}\right]}
\eeq
where $h^{(i)}_{k_{\perp},k}(\eta_R) $ are given by Eqs. (\ref{modefunctions}).

One can carry out the same type of computations in the $L$-wedge of the space-time
arriving to the analogous normalized mode functions in the $L$ section of the $\Sigma$ surface.

\subsubsection{Constructing the vacuum state inside of the lightcone}

Following \cite{Fulling:1977zs} one can define similarly to what we did in the Minkowski case, a new set of mode functions 
that are analytic over the whole Cauchy surface, $\Sigma$. To do that, we introduce the
following normalized functions,
\bea
h^{(i)}_{k_\perp,k}  =
 {1\over{\sqrt{2 \sinh (\pi k /H_{2d})}}} 
\left[e^{\pi k /2H_{2d}}
% \phi^{R,i}(\eta_R,r_R,k,k_{\perp})
\big(h^{(i)}_{k_{\perp},k}(\eta_R)  e^{-i k r_R}\big)
 + e^{-\pi k /2H_{2d}} 
% \overline{\phi^{L,i}(\eta_L,r_L,k,k_{\perp})}
\big((h^{(i)}_{k_{\perp},-k}(\eta_L))^\ast  e^{-i k r_L}\big)
\right] \nonumber
\eea
where $i=1,2$ run over the 2 independent solutions previously found in each %Rindler
$(dS_2)_{R,L}$ wedges.

One can find the form of the vacuum state inside of the light cone, in region M,
by the following analytic continuations of the coordinates,
\begin{eqnarray}
r_R = r_M -  i {{\pi}\over {2H_{2d}}},\quad %\\
\eta_R = -\eta_M -  i {{\pi}\over {2H_{2d}}}~,
\end{eqnarray}
as well as the analogous one for the $L$ coordinates.

Performing this analytic continuation we arrive at
\beq
\phi(\eta_M,r_M,x_{\perp}) = \int{dk \sum_{k_{\perp},i} \left[{1\over {(2 \pi)^{3/2} }}  \tilde a_{k_{\perp},k,i} ~f^{(i)}_{k_{\perp},k}(\eta_M) e^{-i k r_M}  e^{i k_{\perp} x_{\perp}}  + \text{h.c}\right]}~,
\eeq
where
\begin{eqnarray}
f^{(1)} _{k_{\perp},k}(\eta_M)&=& {1\over {\sqrt{2k}}}  {{e^{\pi k /2H_{2d}} }
\over{\sqrt{2 \sinh (\pi k /H_{2d})}}}   N(k,k_{\perp})~\tilde f^{(1)} _{k_{\perp},k}(\eta_M)\\
f^{(2)}_{k_{\perp},k}(\eta_M) &=&  {1\over {\sqrt{2k}}}  {{e^{\pi k /2H_{2d}} }
\over{\sqrt{2 \sinh (\pi k /H_{2d})}}} \left( L(k,k_{\perp}) ~\tilde f^{(1)} _{k_{\perp},k}(\eta_M)
+ e^{-\pi k/H_{2d} }\tilde f^{(2)}_{k_{\perp},k}(\eta_M)\right)
\end{eqnarray}
where we have defined,
\begin{equation}
\tilde f^{(1)} _{k_{\perp},k}(\eta_M)= e^{-ik \eta_M} F\left[-\nu, \nu+1,1-\mu,{{1+\xi_i}\over 2}\right],
\end{equation}
\begin{equation}
\tilde f^{(2)} _{k_{\perp},k}(\eta_M)=  e^{ik \eta_M}F\left[-\nu, \nu+1,1+\mu,{{1+\xi_i}\over 2}\right]~,
\end{equation}
with
\begin{equation}
\xi_i = \coth (H_{2d} \eta_M)~~~;~~~
\mu = i\left( {k\over H_{2d}}\right)~~~;~~~
\nu (\nu +1) = - \left({{k_{\perp}}\over H_{2d}}\right)^2.
\end{equation}\
and where we have simplified the notation by denoting, $F[a,b,c,x]={}_2F_1[a,b,c,x]$.

As we explained in the previous section of this Appendix doing this analytic continuation to
our metric bring us to the other patch of $dS_2$, the one inside of the
lightcone, namely,
\begin{equation}
ds^2_M = {1\over {\sinh^2 (-H_{2d} \eta_M)}} (-d\eta_M^2 +  dr_M^2).
\end{equation}
On the other hand, this metric has the same asymtotic behavior in the $\eta_M \rightarrow  -\infty$ 
limit as the $2d$ part of our anisotropic de Sitter (Kasner-de Sitter) metric so we can take the analytic continuation of our
vacuum as the right initial conditions for the mode functions inside of the decompatification
bubble. This is what we do in the main part of the text.

 \bibliography{anisotropy}

\end{document}